\documentclass[
amsmath,
prd,nofootinbib,floatfix,11pt,
]{revtex4}

\newcommand\beq{\begin{eqnarray}}
\newcommand\eeq{\end{eqnarray}}
\def\lsim{\mathrel{\rlap{\lower4pt\hbox{$\sim$}}
    \raise1pt\hbox{$<$}}}                
\def\gsim{\mathrel{\rlap{\lower4pt\hbox{$\sim$}}
    \raise1pt\hbox{$>$}}}            

\allowdisplaybreaks
\usepackage{graphicx}
\usepackage{setspace}

\begin{document}
\renewcommand{\theequation}{\arabic{section}.\arabic{equation}}
\renewcommand{\thefigure}{\arabic{section}.\arabic{figure}}
\renewcommand{\thetable}{\arabic{section}.\arabic{table}}

\title{\Large \baselineskip=20pt 
Interference of Higgs diphoton signal and background\\
in production with a jet at the LHC}

\author{Stephen P.~Martin}
\affiliation{
{\it Department of Physics, Northern Illinois University, DeKalb IL 60115}  and \\
{\it Fermi National Accelerator Laboratory, P.O. Box 500, Batavia IL 60510}
}

\begin{abstract}\normalsize \baselineskip=14pt 
The Higgs mass determination from diphoton events at the LHC can be affected by interference 
between the Higgs resonant and continuum background amplitudes with the same
initial and final states. For the leading order gluon fusion process, 
this shift was previously found to exceed 100 MeV, with some dependence 
on the diphoton mass resolution and the methods used to extract and fit
the peak from data. In this paper, I consider the mass shift for the process 
$pp \rightarrow j\gamma\gamma$ that includes an additional central jet in the final state.
For cuts on the transverse momentum of the jet of 25 GeV or more, the 
diphoton Higgs peak mass shift due to interference is found to be very small, due in part to 
less interference for the gluon-gluon initiated subprocess, and in part to a cancellation 
between it and the quark-gluon initiated subprocess.  
\end{abstract}


\maketitle


\baselineskip=15.4pt

\setcounter{footnote}{1}
\setcounter{figure}{0}
\setcounter{table}{0}

\section{Introduction\label{sec:intro}}
\setcounter{equation}{0}
\setcounter{figure}{0}
\setcounter{table}{0}
\setcounter{footnote}{1}

The ATLAS and CMS detector collaborations at the LHC have recently 
established \cite{ATLASHiggs,CMSHiggs,ATLAScombination,CMScombination} the existence of a 
resonance whose properties are consistent with those of the minimal 
Standard Model Higgs scalar boson, $H$. 
The properties of this resonance are now
the subject of detailed theoretical and experimental 
investigations to establish its quantum numbers, couplings, and mass.
Given the absence of direct or indirect indications for a non-minimal electroweak symmetry breaking 
mechanism, it will be assumed here that the resonance is indeed $H$. 

The mass determination of $H$ is driven primarily by the invariant mass peaks in the 
$\gamma\gamma$ 
and $ZZ^* \rightarrow \ell^+\ell^-\ell^{\prime +} \ell^{\prime -}$ channels. 
The production of $H$ is mostly due to $gg \rightarrow H$ \cite{Georgi:1977gs}, for which a great effort has been made
to include higher order effects, notably up to
next-to-next-to-leading order (NNLO) in QCD
\cite{Dawson:1990zj,Djouadi:1991tka,Spira:1995rr,deFlorian:1999zd,Ravindran:2002dc,Harlander:2002wh,Glosser:2002gm,Anastasiou:2002yz,Ravindran:2003um,Anastasiou:2005qj,Campbell:2006xx,Campbell:2010cz,vanDeurzen:2013rv,Boughezal:2013uia},
next-to-leading order (NLO) in
electroweak couplings \cite{Aglietti:2004nj,Actis:2008ug,Anastasiou:2008tj},
and next-to-next-to-leading logs in soft gluon resummation
\cite{Catani:2003zt,Bozzi:2005wk,deFlorian:2011xf}. These contributions are reviewed in
\cite{deFlorian:2009hc,Dittmaier:2011ti,Dittmaier:2012vm,Anastasiou:2012hx}.
However, because 
the mass measurement comes from invariant mass distributions, for the most part 
it does not depend directly on the 
details of the Higgs production, including the significant remaining uncertainty on the total rate
and the presence of additional hadrons.
The best experimental values for the mass
combining the $\gamma\gamma$ and $ZZ^*$ channels at this writing are 
\beq
M_H &=& 125.5 \pm 0.2\,  {}^{+0.5}_{-0.6} \,\>{\rm GeV}\qquad\mbox{(ATLAS, \cite{ATLAScombination})},
\\
M_H &=& 125.8 \pm 0.4  \pm 0.4 \,\>{\rm GeV}\qquad\mbox{(CMS, \cite{CMScombination})}.
\eeq
In each case, the first uncertainty is statistical and the second is systematic.
In the future, one may hope to achieve much more precise values, 
given more statistics and reduced systematic uncertainties. Even now, it is worth accounting for effects on the mass 
determination of order 0.1 GeV, since this is the last digit being reported by the experimental collaborations.

One of the issues that may need to be confronted in a precision determination of 
$M_H$ is the effect of the interference 
between resonant Higgs production amplitudes 
and the continuum (non-Higgs-mediated) amplitudes with the same initial and final states. The 
interference effect can produce small 
shifts in the invariant mass distributions, which are in principle observable because they differ for 
different parton-level processes.
In particular, for the diphoton channel 
the interference effect is not completely negligible because of the relatively large continuum 
amplitude 
(one-loop order) compared to the Higgs-mediated amplitude (which is two-loop order; there are no renormalizable couplings of 
the neutral $H$ to $\gamma\gamma$ or to $gg$). In ref.~\cite{hint}, it was shown that in the leading order parton-level 
process $gg \rightarrow \gamma\gamma$, interference effects can shift the position of the Higgs diphoton invariant mass peak 
lower by over 100 MeV compared to where it would be ignoring the interference. Since the latter 
corresponds to what should be obtained in the $ZZ$ and vector boson fusion channels, which will not 
have such a significant 
interference effect, this shift is observable. The magnitude of the shift will depend on the method 
used to fit to the diphoton peak, and will 
also be greatly affected by higher order corrections and by cuts and kinematic-dependent detector 
efficiencies.

In general, the diphoton mass lineshape in proton-proton collisions can be written in terms of the invariant mass of the 
diphoton pair, $\sqrt{h} \equiv M_{\gamma\gamma}$,
as the sum of a continuum plus a 
Breit-Wigner peak multiplied by functions that are approximately symmetric and 
antisymmetric about the Higgs pole mass:
\beq
\frac{d\sigma_{pp\rightarrow \gamma\gamma + X}}{d(\sqrt{h})} &=& C(h) +
\frac{1}{D(h)}\left [P(h) + (h - M_H^2) I(h) \right]
.
\label{eq:genform}
\eeq
Here, $C(h)$, $P(h)$, and $I(h)$ are smooth functions of $h$ 
near the resonance, and
\beq
D(h) \equiv (h - M_H^2)^2 + M_H^2 \Gamma^2_H ,
\eeq
where $M_H$ is the Breit-Wigner mass of the Higgs from the renormalized propagator, and $\Gamma_H$ is the Higgs total decay 
width.
The function $C(h)$ arises from the continuum involving Feynman diagrams that 
do not include the Higgs boson. It falls smoothly with $h$, and is determined by the experimental collaborations by sideband
analyses,
fitting to data 
away from the diphoton peak. Because this is most accurately determined experimentally, it will not be considered as an object 
of theoretical computation here.
The function $P(h)$ arises mostly from the pure Higgs resonance diagrams squared, 
with a small contribution from the interference.
Almost all previous studies of the Higgs diphoton signal have relied on the 
narrow width approximation in which $1/D(h) \approx \pi \delta(h - M_H^2) /M_H \Gamma_H$, 
and one evaluates $H+X$ production separately from the on-shell decays of $H$, including the
diphoton decay 
\cite{Ellis:1975ap,Shifman:1979eb,Gunion:1985dj,Ellis:1987xu,Gunion:1987ke,Djouadi:1997yw}. 
In that approximation, the function $I(h)$ does not appear.
In general, the function $I(h)$ arises only from the interference terms between Higgs resonant 
and continuum amplitudes. Its importance is that it 
gives rise to a shift in the 
diphoton mass distribution peak away from $M_H$, since the corresponding contribution to the 
cross-section is odd in $\sqrt{h} - M_H$. The sign of the shift in the diphoton mass peak, 
compared to its position if interference were neglected, is the same as the sign of $I(h)$.  
The magnitude of the mass shift depends on the relative sizes 
of $I(h)$ and $P(h)$ with kinematic cuts (to be evaluated numerically below) 
and detector effects including the diphoton mass resolution.
 
In contrast, the effect of interference on the total cross-section is 
very small at leading order \cite{Dicus:1987fk,Dixon:2003yb}, while at next-to-leading order 
there is a reduction of a few per 
cent \cite{Dixon:2003yb} due to the imaginary part of the 2-loop continuum amplitude 
$gg \rightarrow \gamma\gamma$ from light quark loops
\cite{Bern:2001df}. Other studies of the effects of interference between signal 
and background in Higgs production in different 
contexts can be found in 
refs.~\cite{Glover:1988fe,Dixon:2008xc,Campbell:2011cu,Passarino:2012ri,Kauer:2012hd}.

The leading order shift in the Higgs mass peak due to interference should be investigated with 
a full NLO calculation, at least. As a 
precursor to this, in the present paper I will investigate the 
interference between signal and background for processes 
contributing to diphoton production with an additional central jet requirement 
imposed on the final state, $pp \rightarrow 
j\gamma\gamma$. The parton-level processes $Qg \rightarrow Q\gamma\gamma$ 
and $\overline Qg \rightarrow \overline Q\gamma\gamma$ and $Q \overline Q \rightarrow g \gamma\gamma$ 
are suppressed by relatively small quark parton distribution functions, but this is counteracted in 
part by the fact that the continuum amplitudes are tree-level, providing for a stronger interference 
with the Higgs resonant amplitudes, compared to the non-interference contributions. These processes
have recently been investigated in \cite{deFlorianHj}, where it is found that the 
diphoton mass distribution shift is in the 
opposite direction to the leading order $gg \rightarrow \gamma\gamma$ shift. I find agreement with 
their result, and in the present paper will include also the $ gg \rightarrow g\gamma\gamma$
process, which has a mass shift with the same sign as the shift from  $gg \rightarrow \gamma\gamma$. 

Previous investigations 
\cite{Abdullin:1998er,deFlorian:1999zd,Zmushko:2002fva,Demidov:2004qt,Brein:2007da} 
of the $pp \rightarrow jH$ signal for the LHC have considered 
a cut on $p_T^j$ of 30 GeV or higher. In the present paper this cut will 
be varied to both much larger and much smaller values. In the limit that the 
$p_T^j$ cut on the final-state jet is taken to be very small 
(certainly for less than 15 GeV or so), the results are
clearly unphysical, as the real emission of a soft jet is subject 
to infrared log divergences that should be regularized 
and canceled against 
those coming from virtual corrections to the leading-order process 
$gg \rightarrow \gamma\gamma$ in a full NLO calculation. 
Nevertheless, I will include below the experimentally unrealistic case of very 
low $p_T^j$ cuts even below 1 GeV, since this 
provides a check; the result for the mass shift due to interference in this case approaches that 
for the leading-order process, as the calculated
production is dominated by the leading order subdiagrams $gg\rightarrow \gamma\gamma$ with a soft 
gluon emission attached to them.

The rest of this paper is organized as follows. In section \ref{sec:H}, the situation for 
the leading order process without an 
extra jet is reviewed, following ref.~\cite{hint}, and including numerical results 
for the same cuts on the photons as will be 
imposed later on the process with an additional jet. Section \ref{sec:jH} provides 
analytical formulas for the pure Higgs and 
interference contributions to $pp \rightarrow j\gamma\gamma$. 
Numerical results are then discussed in section \ref{sec:num}. 
Section \ref{sec:outlook} contains some summarizing remarks.

\section{Higgs interference in $pp \rightarrow \gamma\gamma$\label{sec:H}}
\setcounter{equation}{0}
\setcounter{figure}{0}
\setcounter{table}{0}
\setcounter{footnote}{1}

The leading order diphoton production cross-section relevant to Higgs production and interference can be written as
\beq
\frac{d\sigma_{pp\rightarrow \gamma\gamma}}{d(\sqrt{h})} &=&
\frac{1}{128\pi\sqrt{h} D(h)} 
\int_{\ln\sqrt{\tau}}^{-\ln\sqrt{\tau}} \frac{dy}{s}
\, g(\sqrt{\tau} e^y, \mu_F^2) g(\sqrt{\tau} e^{-y}, \mu_F^2) 
\, 
\int_{-1}^1 dz\>\Theta(h,y,z) \> N(h,z),
\phantom{xxxx}
\label{eq:dsigmadhLO}
\eeq
where $\tau = h/s$, 
with $\sqrt{s}$ the fixed total energy of the $pp$ collisions at the LHC,
$g(x,\mu_F^2)$ is the gluon parton distribution function,
$y$ is the longitudinal rapidity of the partonic center-of-momentum frame,
$z$ is the cosine of the photon scattering angle with respect to the beam axis, and
$\Theta(h,y,z)$ represents the effects of kinematic cuts. The resonant and 
interference contributions to $N(h,z)$ are $N_H + N_{\rm int,Re} + N_{\rm int,Im}$, with 
\beq
N_H &=& h^2 |C_g C_\gamma|^2 /4,
\\
N_{\rm int,Re} &=& -(h - M_H^2) h\, {\rm Re} [C_g C_\gamma A^*_{gg\gamma\gamma}] ,
\\
N_{\rm int,Im} &=& -M_H \Gamma_H h\, {\rm Im} [C_g C_\gamma A^*_{gg\gamma\gamma}] .
\eeq
Here, the effective Higgs coupling to gluons, in the limit of a very heavy 
top quark and other quarks massless, is parameterized by
\beq
C_g = \frac{\alpha_S}{3 \pi v},
\label{eq:defCg}
\eeq
using a normalization where $v\approx 246$ GeV is the Higgs expectation value. 
This $M_t \rightarrow \infty$ effective theory for the 
Higgs interactions with gluons (both $Hgg$ and $Hggg$)  
is a good approximation \cite{Shifman:1979eb,Dawson:1990zj,Djouadi:1991tka,Kauffman} 
for the realistic case 
(with $M_H \sim 125$ GeV and $M_t = 173$ GeV) for transverse momenta less than $M_t$, and 
will be used throughout this paper. 
The Higgs interaction with photons is instead treated using the complete one-loop expression:
\beq
C_{\gamma} &=& -\frac{\alpha h}{4\pi v} \biggl [
F_1(4 m_W^2/h) + 
\sum_{f = t,b,c,\tau} N_{c}^f e_f^2 F_{1/2} (4 m_f^2/h) \biggr ],
\label{eq:defCgamma}
\eeq
where $N_c^f=3$ (1) for $f=$ quarks (leptons) with electric charge $e_f$ and mass $m_f$, and
\beq
F_1(x) &=& 2 + 3 x [1 + (2-x) f(x)],
\\
F_{1/2} (x) &=& -2 x [1 + (1-x) f(x)],
\\
f(x) &=& \Biggl \{ \begin{array}{ll}
[\arcsin (\sqrt{1/x})]^2, & x\geq 1 \quad \mbox{(for $t,W$)},
\\
-\frac{1}{4} \left [ \ln \left (\frac{1 + \sqrt{1-x}}{1 - \sqrt{1-x}}\right ) 
- i \pi \right ]^2,\phantom{xxx}
& x\leq 1 \quad \mbox{(for $b,c,\tau$)}.
\end{array}
\eeq
(The effective Higgs couplings used in ref.~\cite{hint} are related to these definitions 
by $A_{\gamma\gamma H} = C_\gamma$ and $A_{ggH} = h C_g/2$, and the variable $\hat s$ 
there is the same as $h$ here.) 
For the
continuum amplitude contribution \cite{Karplus:1950zz,Costantini:1971cj,Combridge:1980sx}, the heavy top and massless $u,d,c,s,b$ approximation
is also used here, leading to
\beq
A_{gg\gamma\gamma} = \frac{22}{9} \alpha_S \alpha \left \{
z \ln \left (\frac{1+z}{1-z}\right ) - \frac{1+z^2}{4} 
\left [ \ln^2 \left (\frac{1+z}{1-z}\right ) + \pi^2 \right ] \right \} .
\label{eq:Aggaa}
\eeq
The numerical effect of including a finite top mass and non-zero bottom mass is not 
very large for the interference effect as it applies to the diphoton mass shift.

For $pp \rightarrow \gamma\gamma$, the cuts on the transverse momenta and pseudo-rapidity 
of the photons are
\beq
&&
p_{T\gamma} > p_{T\gamma}^{\rm cut} = 40\>{\rm GeV},
\\
&& 
|\eta_{\gamma}| < \eta_\gamma^{\rm cut} = 2.5.
\eeq
These cuts are implemented in the numerical integration of this section (with no extra jet)
simply by imposing the restrictions that $|y|<\eta_\gamma^{\rm cut}$ and 
that $|z|$ is less than both  $\sqrt{1 - 4 (p_{T\gamma}^{\rm cut})^2/h}$ and 
$\tanh(\eta_\gamma^{\rm cut} - |y|)$. The results below therefore differ from 
ref.~\cite{hint}, where these cuts on the photons were mentioned but 
not directly applied. The impact of this is to 
reduce the mass shift due to the interference somewhat.

\begin{figure}[!t]
\begin{minipage}[]{0.49\linewidth}
\includegraphics[width=0.9\linewidth,angle=0]{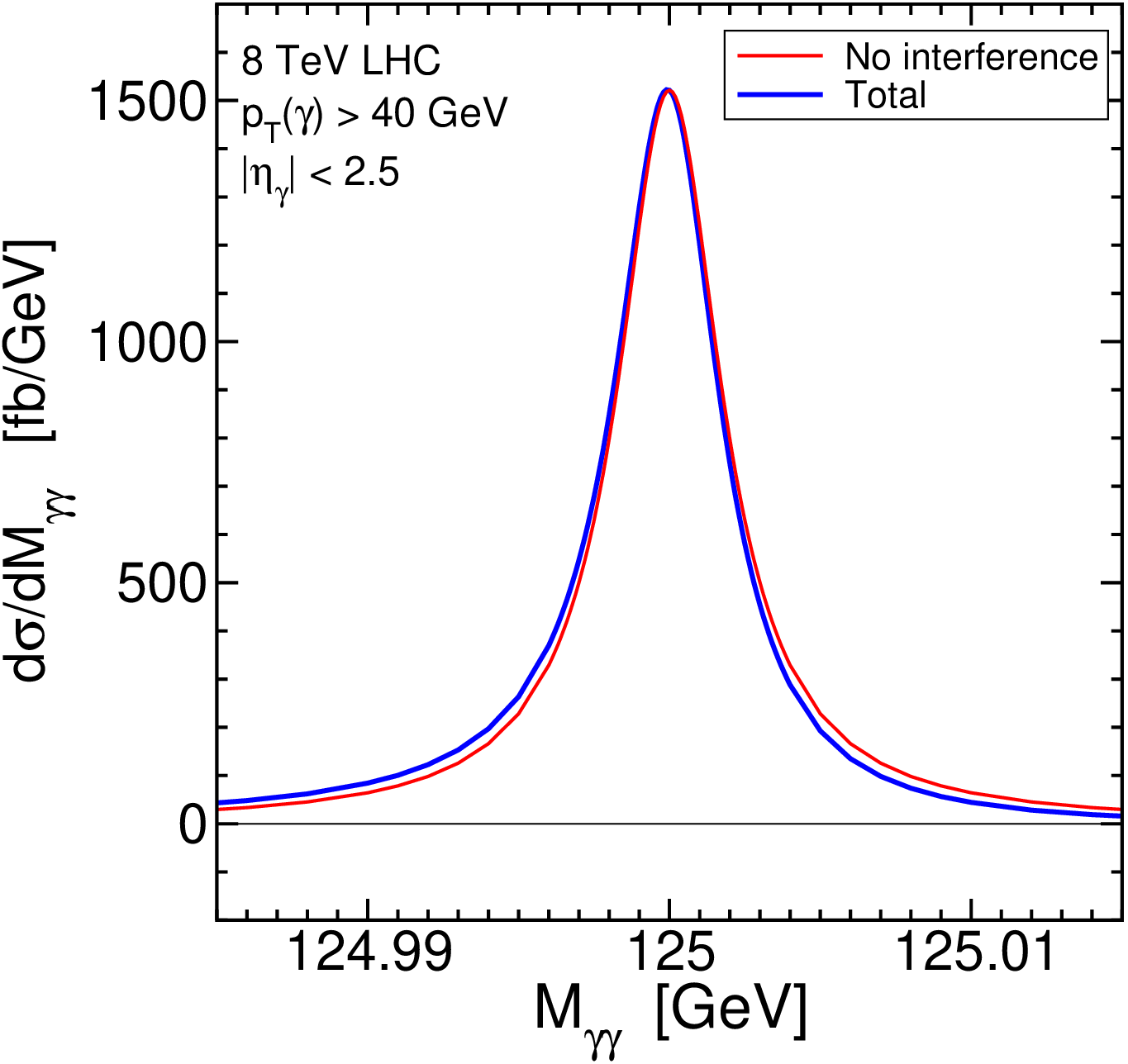}
\end{minipage}
\begin{minipage}[]{0.49\linewidth}
\begin{flushright}
\includegraphics[width=0.9\linewidth,angle=0]{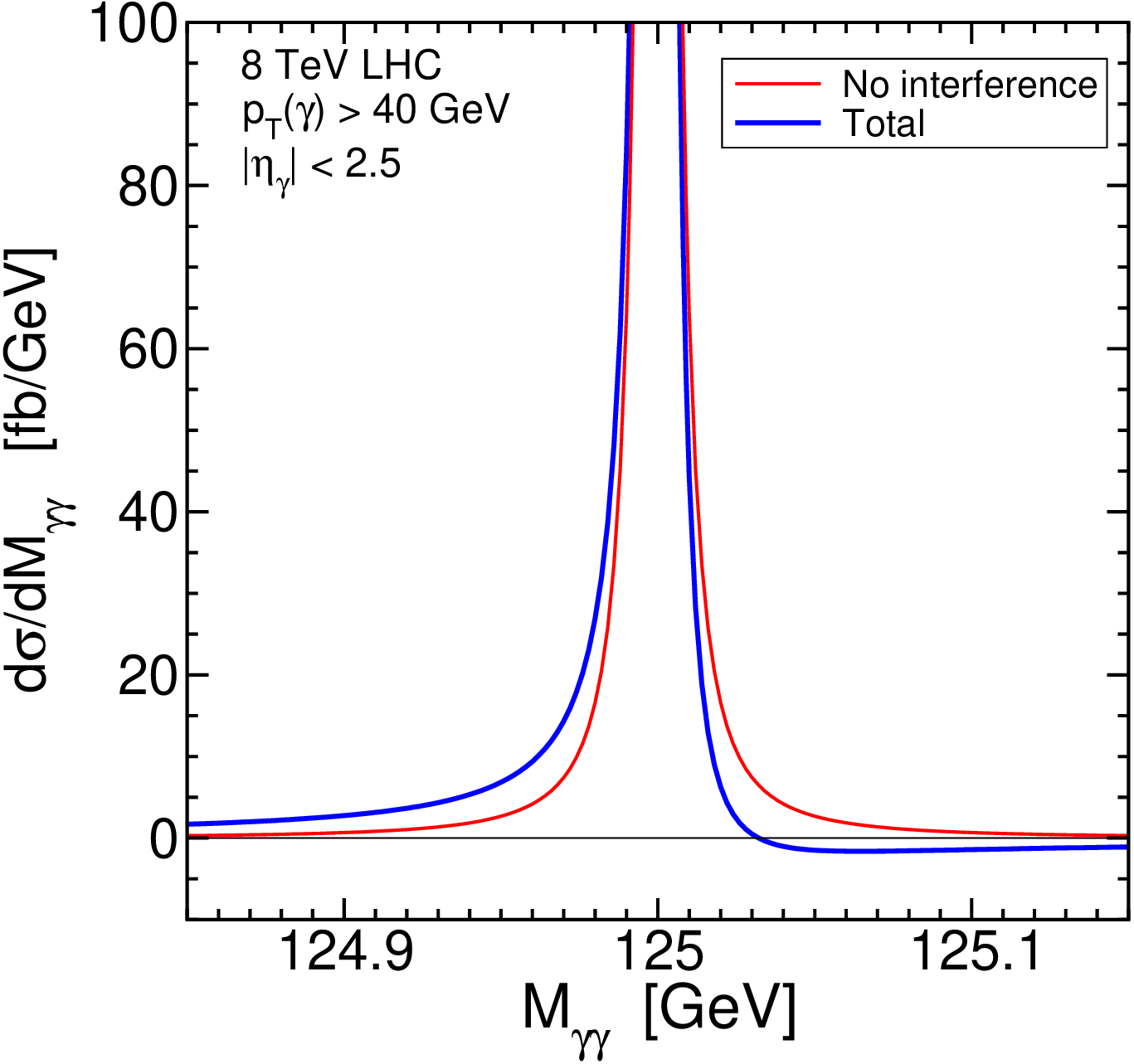}
\end{flushright}
\end{minipage}
\begin{minipage}[]{0.49\linewidth}
\includegraphics[width=0.9\linewidth,angle=0]{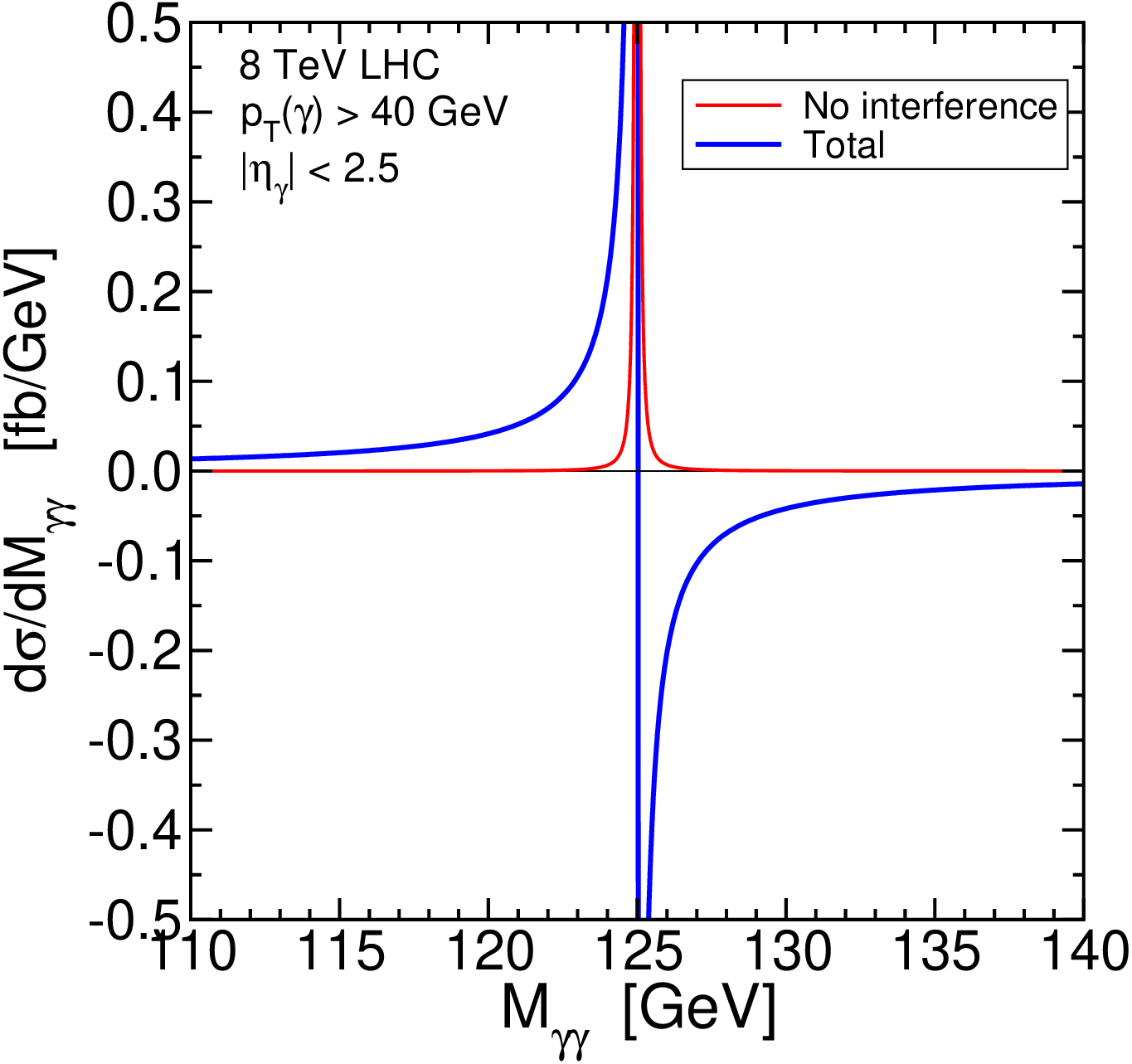}
\end{minipage}
\begin{minipage}[]{0.49\linewidth}
\begin{minipage}[]{0.04\linewidth}\end{minipage}
\begin{minipage}[]{0.90\linewidth}
\caption{\label{fig:unsmearedLO}
The diphoton invariant mass distribution, with and without interference included, before including any 
experimental resolution effects, computed for 
$pp \rightarrow \gamma\gamma$ at leading order as described in section \ref{sec:H} 
from the partonic process $gg \rightarrow \gamma\gamma$,
for 8 TeV $pp$ collisions at the LHC, with $p_T > 40$ GeV and $|\eta| < 2.5$ for the photons. 
Here $M_{\gamma\gamma} = \sqrt{h}$ is the diphoton mass.
The three panels show exactly the same results but with different scales on the axes.
} \end{minipage}\end{minipage}
\end{figure}

For purposes of illustration, I take
$M_H=$ 125 GeV and $\Gamma_G = 4.2$ MeV. The parameter $C_\gamma$ is evaluated using 
$m_t = 168.2$ GeV, $m_b = 2.78$ GeV, $m_c = 0.72$ GeV, $m_\tau = 1.744$ GeV, and $\alpha = 1/127.5$.
Also, to facilitate comparison with an eventual NLO calculation, I have used MSTW 2008 NLO 
\cite{Martin:2009iq}
parton distribution functions with factorization scale $\mu_F = M_H$, and evaluated 
the corresponding strong coupling at the same renormalization scale $\mu_R = M_H$; explicitly this is
$\alpha_S(M_H) = 0.114629$. The unsmeared diphoton lineshape is shown in Figure \ref{fig:unsmearedLO}. 
For $\sqrt{h}$ very close to $M_H$, the lineshapes are nearly indistinguishable, but for
$|\sqrt{h} - M_H| \gsim 50$ MeV, the magnitude of the interference term is much larger than the
pure resonance contribution, due to the long tails of the square root of the Breit-Wigner lineshape.
The effect of the interference is to produce slightly more events below $M_H$ than above $M_H$, because 
the function $I(h)$ in eq.~(\ref{eq:genform}) is negative near $\sqrt{h}= M_H$.

\begin{figure}[!t]
\begin{minipage}[]{0.49\linewidth}
\includegraphics[width=\linewidth,angle=0]{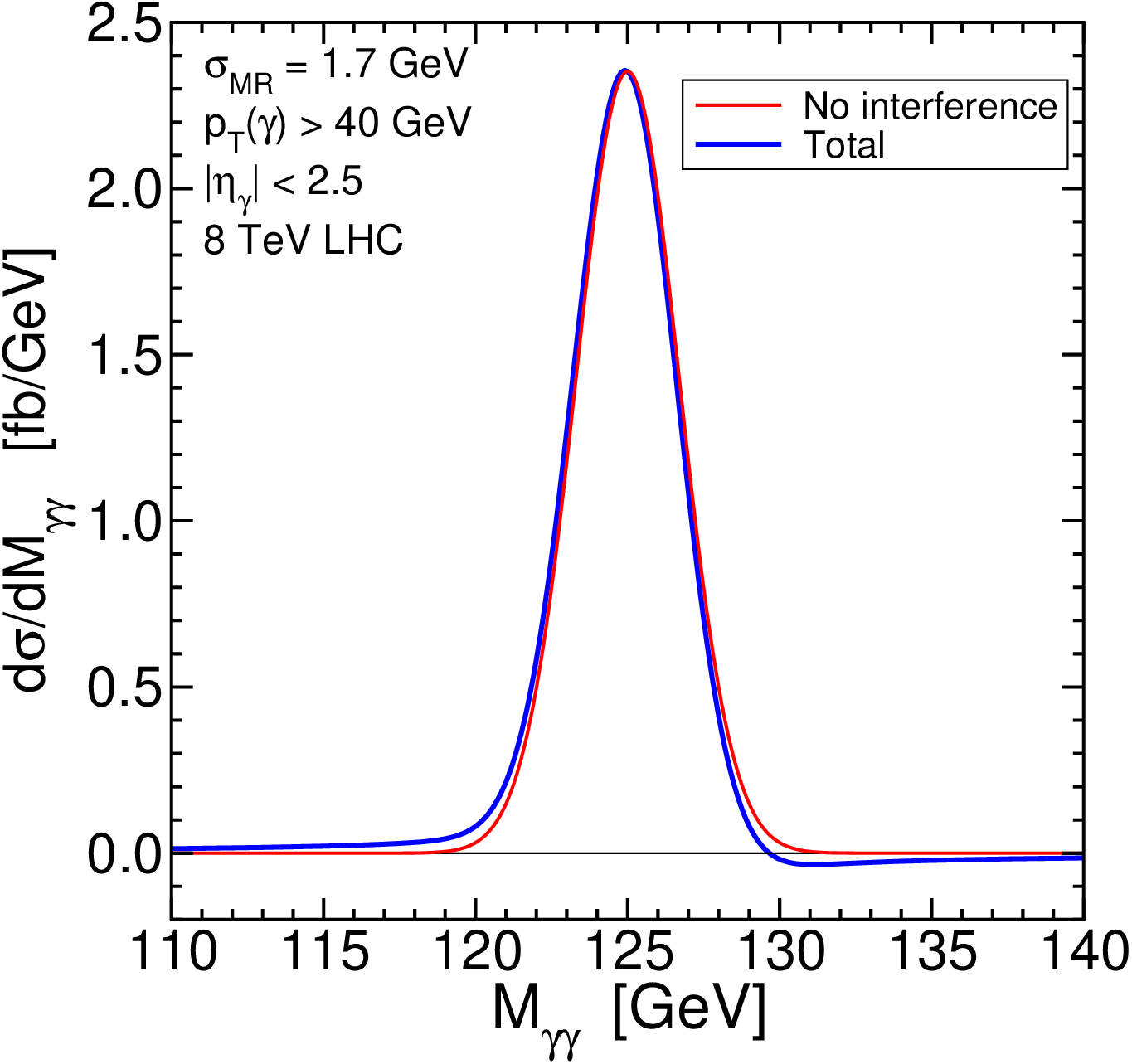}
\end{minipage}
\begin{minipage}[]{0.49\linewidth}
\begin{flushright}
\includegraphics[width=\linewidth,angle=0]{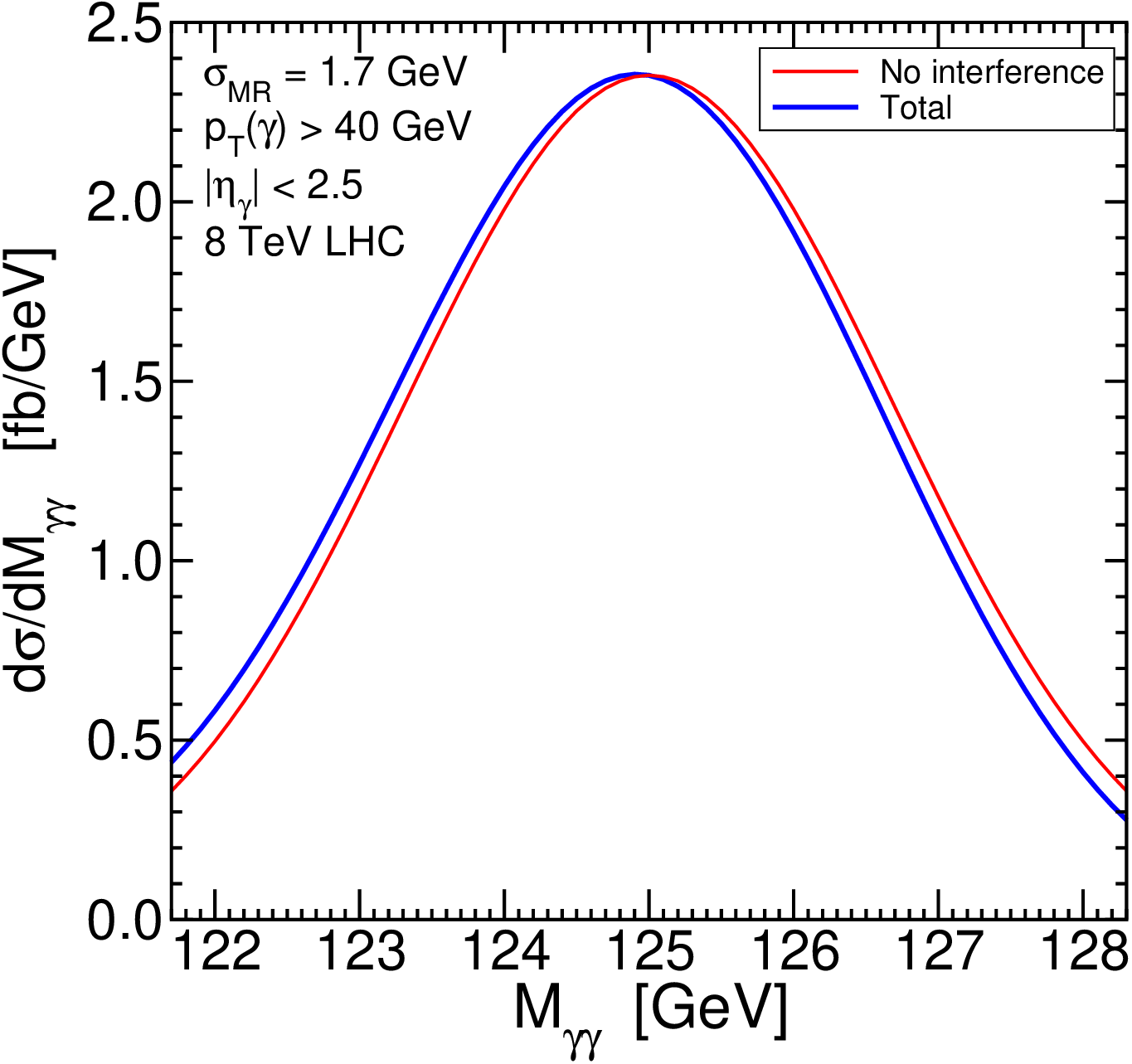}
\end{flushright}
\end{minipage}
\caption{\label{fig:smearedLO}
The diphoton invariant mass distribution for $pp\rightarrow \gamma\gamma$ at leading order, 
with and without interference included, as in Figure
\ref{fig:unsmearedLO}, but now including the effects of 
a Gaussian mass resolution with $\sigma_{\rm MR} = 1.7$ GeV. 
The two panels show the same results with different scales on the axes.}
\end{figure}

The effects of detector resolution are complicated, depending on the location and type of
interaction of 
photons in the detector. For simplicity, I assume a Gaussian invariant mass resolution, with mass resolution widths $\sigma_{\rm MR}$. For a typical case $\sigma_{\rm MR} = 1.7$ GeV, the diphoton lineshape after this Gaussian smearing is shown in Figure \ref{fig:smearedLO}.
After Gaussian smearing there 
remains a potentially detectable shift in the diphoton mass distribution.

\begin{figure}[!t]
\begin{minipage}[]{0.49\linewidth}
\begin{flushright}
\includegraphics[width=\linewidth,angle=0]{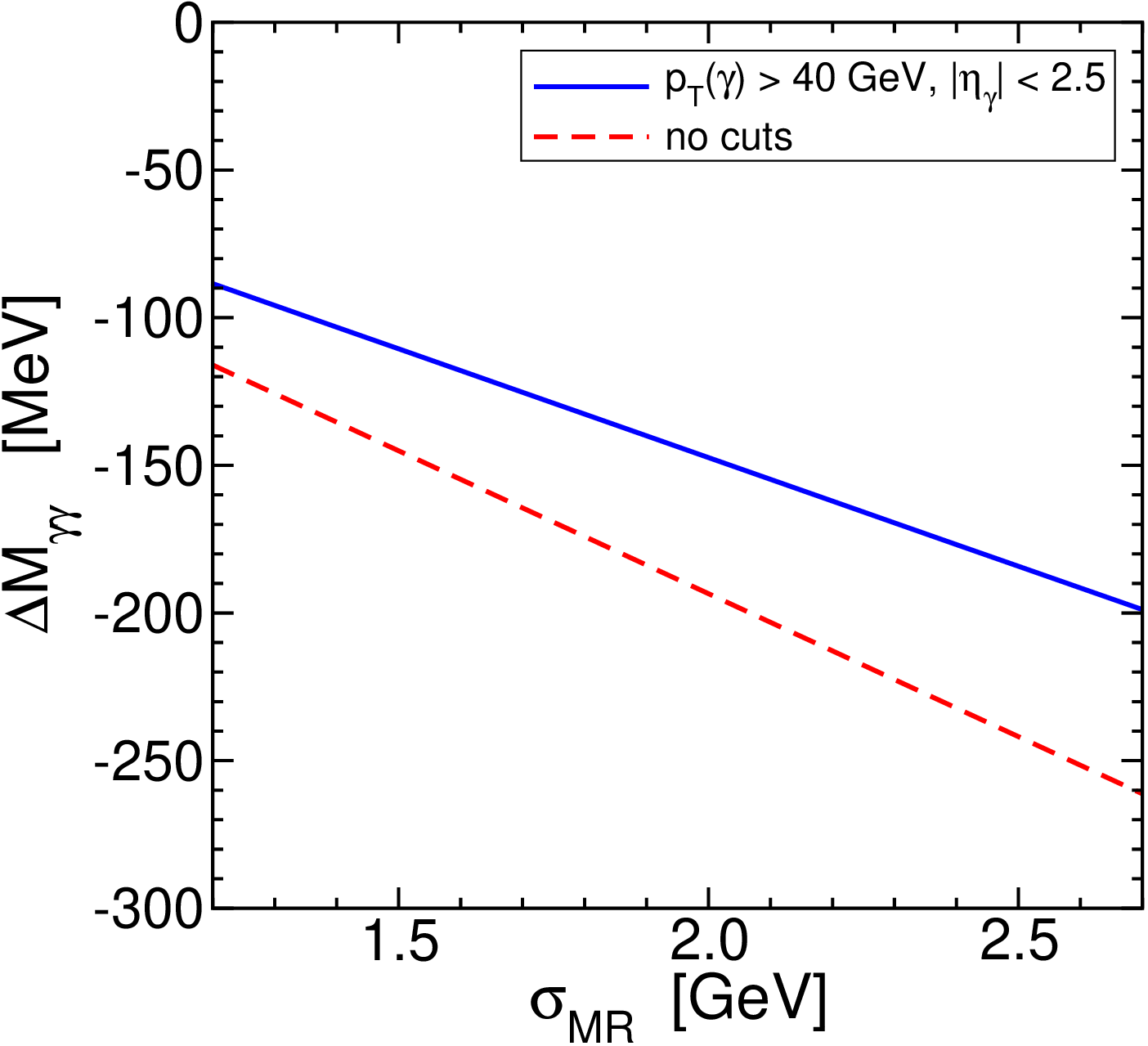}
\end{flushright}
\end{minipage}
\begin{minipage}[]{0.49\linewidth}
\begin{minipage}[]{0.04\linewidth}\end{minipage}
\begin{minipage}[]{0.90\linewidth}
\caption{\label{fig:shiftLO}
The shift in the invariant mass distribution due to the interference effect, 
$\Delta M_{\gamma\gamma} \equiv M_{\gamma\gamma}^{\rm peak} - M_H$, 
for $pp \rightarrow \gamma\gamma$ at 
leading order, computed by a least-squares fit of the lineshape to a Gaussian with the 
same width $\sigma_{\rm MR}$ used to model the mass resolution. The solid line includes cuts
$p_T > 40$ GeV and $|\eta| < 2.5$ on the photons, and the dashed line is what would be obtained 
without these cuts.}
\end{minipage}\end{minipage}
\end{figure}

The magnitude of this shift will depend on the methods used by the experimental collaborations to fit
to the lineshape, in particular the background. In \cite{hint}, one measure of this shift 
was described, but a simpler and better method is to simply do a 
least-squares fit of the lineshapes 
with and without interference to a Gaussian with the same width 
$\sigma_{\rm MR}$ as was used to model the 
mass resolution. For the purely resonant contribution without interference included, the peak 
of the distribution is at $\sqrt{h} = M_H$ to very high accuracy.
In the following, the difference between the centers of the Gaussian fits 
with and without interference included will be 
called $\Delta M_{\gamma\gamma} \equiv M_{\gamma\gamma}^{\rm peak} - 
M_H$. The fit is performed over a range 
of $\sqrt{h}$ from 115 GeV to 
135 GeV, but the results are not very sensitive to this particular choice. (Even a 
range 120 to 130 GeV gives nearly the same results, except when $\sigma_{\rm MR}$ is larger than about 
2.5 GeV.) The magnitude of the shift by this measure is shown in Figure \ref{fig:shiftLO},
for varying $\sigma_{\rm MR}$ used for both the smearing and the fit.
The magnitude of the shift according to this measure actually increases nearly linearly 
with increasing mass resolution width $\sigma_{\rm MR}$.
For a typical average value $\sigma_{\rm MR} = 1.7$ GeV, the shift is about 
$\Delta M_{\gamma\gamma} = -125$ MeV after cuts; it would be about $-165$ MeV before the 
photon $p_T$ and $\eta$ cuts. This is because the continuum amplitude has larger support 
at small scattering angles ($z$ near $\pm 1$),
due to the logarithms in eq.~(\ref{eq:Aggaa}), while the Higgs resonant amplitude is isotropic 
in the partonic center-of-momentum frame. 

The previous results were made with the somewhat arbitrary fixed scale choices $\mu_R = \mu_F = M_H$.
However, variations in these scale choices for the strong coupling and the parton distribution 
functions tend to nearly cancel out of $\Delta M_{\gamma\gamma}$, because they enter into the 
interference term and the pure resonance term in the same way. The choice made here
of using the NLO rather than the LO $\alpha_S(M_H)$ makes the computed total cross sections 
smaller by
about 33\%. However, since $I(h)$ and $P(h)$ are both proportional to $\alpha_S^2$, this 
dependence very nearly cancels out of the prediction for $\Delta M_{\gamma\gamma}$.

\section{Higgs interference in $pp \rightarrow j\gamma\gamma$\label{sec:jH}}
\setcounter{equation}{0}
\setcounter{figure}{0}
\setcounter{table}{0}
\setcounter{footnote}{1}

Now consider the process of Higgs production in association with a jet, in the case where the 
Higgs decays to two photons. Because the relevant parton level processes
$gg \rightarrow g\gamma\gamma$ and $Qg \rightarrow Q\gamma\gamma$ 
and $\overline Qg \rightarrow \overline Q\gamma\gamma$ and $Q \overline Q \rightarrow g \gamma\gamma$ have different initial and final states than the 
$gg\rightarrow \gamma\gamma$ case studied in the previous section and in \cite{hint}, it will 
be no surprise that the interference effect on the mass shift will be different when an extra 
jet is required by the selection. In fact, the processes involving quarks have continuum amplitudes
already at tree-level, which provides for a stronger interference with the Higgs resonant 
amplitudes, compared to the Higgs-only cross sections. However, this effect is mitigated by the 
smaller quark parton distribution functions for the relevant momentum fractions.
 
Let us label the initial state partons by 1,2, and the final state jet 
parton by 3, and the final state photons by 4,5. The corresponding momenta and helicities are 
denoted $(p_i, \lambda_i)$ for 
$i=1,2,3,4,5$. Amplitudes below are evaluated using the spinor helicity formalism following the 
conventions of refs.~\cite{ManganoParke,DixonTASI} for spinor products, and 
using a convention in which momenta and helicities are always outgoing, even for initial-state particles.

The 4-momenta of the 
partons are parameterized in terms of the 
quantities: $\hat s$ (the invariant squared mass of the initial-state partons), 
$h$ (the invariant squared mass of the two photons), 
$\chi$ (related to the scattering angle of the final-state jet parton), 
and $\omega$, $\phi$ (related to the angles of the individual photons in 
the diphoton system rest frame), as follows. In the lab frame,
\beq 
p_1 &=& -\frac{\sqrt{\hat s}}{2} (1,0,0,1),
\\
p_2 &=& -\frac{\sqrt{\hat s}}{2} (1,0,0,-1),
\\
p_3 &=& \frac{\sqrt{\hat s}}{2}(1-h/\hat s) \left (1,\>2\sqrt{\chi(1-\chi)},\> 0,\> 1 - 2 \chi \right ),
\\
p_H &=& \frac{\sqrt{\hat s}}{2}(1-h/\hat s) \left (\frac{\hat s+h}{\hat s-h}, \>-2\sqrt{\chi(1-\chi)},\> 0,\> -1 + 2 \chi\right ) ,
\eeq
where $H$ denotes the Higgs (or diphoton system), with $p_H = p_4 + p_5$. 
Now $(p_H, p_4, p_5)$ are related to $(p_H', p_4', p_5')$ by an appropriate boost, where in the diphoton system rest frame,
\beq
p_H' &=& \sqrt{h} (1,0,0,0),
\\
p_4' &=& \frac{\sqrt{h}}{2} \left (1,\> 2 \sqrt{\omega(1-\omega)}\,\cos\phi,\> 
2 \sqrt{\omega(1-\omega)}\,\sin\phi,\> 1 - 2 \omega \right ) ,
\\
p_5' &=& \frac{\sqrt{h}}{2} \left (1,\> -2 \sqrt{\omega(1-\omega)}\,\cos\phi,\> 
-2 \sqrt{\omega(1-\omega)}\,\sin\phi,\> -1 + 2 \omega \right ).
\eeq
(The boost is not written explicitly here, but is determined by the relationship of $p_H$ and $p_H'$.)
The ranges for the angular variables are $0 \leq \chi,\omega \leq 1$ and $0 \leq \phi < 2\pi$.
Also, define $s_{ij} = (p_i + p_j)^2$. Note that
\beq
s_{12} = (p_1 + p_2)^2 = (p_3+p_4+p_5)^2 = \hat s,\qquad\qquad
s_{45} = (p_4 + p_5)^2 = h .
\eeq

We are interested in the diphoton line shape,
\beq
\frac{d\sigma_{pp\rightarrow j\gamma\gamma}}{d(\sqrt{h})} &=&
\int d\tau 
\int_{\ln\sqrt{\tau}}^{-\ln\sqrt{\tau}} dy
\, f_1(\sqrt{\tau} e^y,\mu_F^2) f_2(\sqrt{\tau} e^{-y},\mu_F^2) 
\, \frac{d\hat\sigma_{12 \rightarrow 345}}{d(\sqrt{h})}
\label{eq:dsigmadh}
\eeq
where $f_{1,2}$ are the
distribution functions for the initial-state partons 1 and 2 (which should be summed over), and now
$\tau = \hat s/s$. 
Including a factor of $1/2$ for identical photons, the parton-level differential cross-section is:
\beq
\frac{d\hat\sigma}{d(\sqrt{h})} &=& \frac{\sqrt{h}}{512 \pi^4 \hat s} (1 - h/\hat s) 
\int_0^1 d\chi \int_0^1 d\omega \int_0^{2\pi} d\phi\>\>  \Theta\> 
{\textstyle \overline{\sum}}\, |{\cal M}|^2 .
\eeq
Here ${\cal M}$ is the reduced matrix element for $12\rightarrow 345$, and
$\overline{\sum}$ denotes the average 
(and sum) over initial (final) state helicities and colors, and
$\Theta(\hat s, h, y, \chi, \omega, \phi)$ represents the effects of kinematic cuts, implemented below at parton level in a numerical integration.

\subsection{$gg \rightarrow g\gamma\gamma$\label{sec:gggAA}}
\setcounter{equation}{0}
\setcounter{figure}{0}
\setcounter{table}{0}
\setcounter{footnote}{1}

Consider first the process 
\beq
g(p_1, \lambda_1, a) \>+\> g(p_2, \lambda_2, b)
&\longrightarrow&
g(p_3, \lambda_3, c)
\>+\> 
\gamma(p_4, \lambda_4) 
\>+\> 
\gamma(p_5, \lambda_5) 
,
\label{eq:gggAA}
\eeq
with the momenta $p_i$ and the polarizations $\lambda_i = \pm$ taken to be outgoing, and
$a,b,c$ are color adjoint labels. The corresponding matrix element can be written as a sum of 
continuum and resonant Higgs-mediated parts:
\beq
{\cal M} ={\cal M}^{\rm cont} + {\cal M}^H .
\label{eq:MgggAAdecompose}
\eeq
For the Higgs-mediated contribution in eq.~(\ref{eq:MgggAAdecompose}), we will treat the 
gluon couplings to the Higgs using the effective theory in which the top quark is taken very heavy, 
$M_t \gg M_H$. Then one finds 
\beq
{\cal M}^H_{\lambda_1\lambda_2\lambda_3\lambda_4\lambda_5} = 
\frac{g_3}{\sqrt{2}} f^{abc} C_g C_{\gamma}
X_{\lambda_1\lambda_2\lambda_3} Y_{\lambda_4\lambda_5}/\left ( h - M_H^2 + i M_H \Gamma_H \right ),
\label{eq:MHgggAA}
\eeq
with spinor-helicity factors:  
\beq
&&
X_{+++} \>=\> -ih^2/\langle 12 \rangle \langle 23 \rangle \langle 31 \rangle,
\qquad\qquad
X_{---} \>=\> ih^2/[12][23][31],
\label{eq:X}
\\
&&
X_{++-} \>=\> i[12]^3/[23][31],
\qquad\qquad\qquad
X_{--+} \>=\> -i\langle 12\rangle^3/\langle 23\rangle\langle 31\rangle, \phantom{xxxx}
\\
&&
X_{+-+} \>=\> i[31]^3/[12][23],
\qquad\qquad\qquad
X_{-+-} \>=\> -i\langle 31\rangle^3/\langle 12\rangle\langle 23\rangle,
\label{eq:XX}
\\
&&
X_{-++} \>=\> i[23]^3/[31][12],
\qquad\qquad\qquad
X_{+--} \>=\> -i\langle 23\rangle^3/\langle 31\rangle\langle 12\rangle,
\label{eq:XXX}
\\
&&
Y_{++} \>=\> [45]/\langle 45 \rangle,
\qquad\qquad\qquad\qquad\>\>\>\>\,
Y_{--} \>=\> \langle 45 \rangle/[45],
\label{eq:Ypp}
\\ &&
Y_{+-} \>=\> Y_{-+} \>=\> 0.
\label{eq:Ypm}
\eeq
Note that these obey $\langle ij \rangle \leftrightarrow [ji]$ when the
helicities are flipped.
The structure constants of the group are normalized so that $f^{abc} f^{abd} = N \delta^{cd}$ with $N=3$ for QCD.

The continuum matrix element in eq.~(\ref{eq:MgggAAdecompose}) can be given in terms
of the one-quark-loop 5-gluon partial amplitudes 
$A^{[1/2]}_{5;1}$ that were obtained by Bern, Dixon and Kosower in \cite{Bern:1993mq}. 
(These are somewhat complicated, and so will not be reproduced explicitly here. Note that 
flipping all of the helicities on $A^{[1/2]}_{5;1}$ can be obtained by replacing 
$\langle ij \rangle \leftrightarrow [ji]$ everywhere. In particular, $\varepsilon(i,j,m,n) = 
[ij]\langle jm \rangle [mn] \langle ni \rangle - 
\langle ij \rangle [jm] \langle mn \rangle [ni]$ changes sign.) 
One finds, for massless quarks $u,d,s,c,b$ circulating around the loop, and 
neglecting the suppressed top-quark contribution:
\beq
{\cal M}^{\rm cont}_{\lambda_1\lambda_2\lambda_3\lambda_4\lambda_5} &=& 
\frac{g_3}{\sqrt{2}} f^{abc} \left (\frac{44}{9} \alpha_S \alpha \right )
A_{{\lambda_1} {\lambda_2}{\lambda_3}{\lambda_4}{\lambda_5}}
,
\eeq
where  \cite{deFlorian:1999tp,Balazs:1999yf}:
\beq
A_{{\lambda_1}{\lambda_2}{\lambda_3}{\lambda_4}{\lambda_5}} &=& 16 \pi^2 \Bigl [
A^{[1/2]}_{5;1}(1,2,3,4,5) + A^{[1/2]}_{5;1}(1,2,3,5,4) + A^{[1/2]}_{5;1}(1,2,4,3,5) + 
\nonumber \\ && 
A^{[1/2]}_{5;1}(1,2,5,3,4) + A^{[1/2]}_{5;1}(1,2,4,5,3) + A^{[1/2]}_{5;1}(1,2,5,4,3) + 
\nonumber \\ &&
A^{[1/2]}_{5;1}(1,4,2,3,5) + A^{[1/2]}_{5;1}(1,4,2,5,3) + A^{[1/2]}_{5;1}(1,4,5,2,3) +
\nonumber \\ && 
A^{[1/2]}_{5;1}(1,5,2,3,4) + A^{[1/2]}_{5;1}(1,5,2,4,3) + A^{[1/2]}_{5;1}(1,5,4,2,3) \Bigr ] .
\eeq
The individual loop amplitudes $A^{[1/2]}_{5;1}$ have infrared divergences, which cancel in the sum. 

The spin and color sum/average for the reaction eq.~(\ref{eq:gggAA}) is
\beq
{\textstyle \overline{\sum}} &\equiv& \frac{1}{4} \sum_{\lambda_1,\lambda_2,\lambda_3,\lambda_4,\lambda_5} \frac{1}{(N^2-1)^2}
\sum_{a,b,c}.
\eeq
Taking into account $Y_{+-}=Y_{-+}=0$ and $\langle ij \rangle [ij] = -s_{ij}$,
it follows that:
\beq
{\textstyle \overline{\sum}}\, |{\cal M}^H|^2 &=&
\frac{3 \pi \alpha_S}{4 D(h)}
|C_g C_{\gamma}|^2 
\left ( \frac{h^4 + \hat s^4 + s_{13}^4 + s_{23}^4}{\hat s s_{13} s_{23}} \right )
\label{eq:MH2}
\\
{\textstyle \overline{\sum}}\, 2 {\rm Re}[{\cal M}^H {\cal M}^{{\rm cont}*} ] &=&
\frac{11\pi\alpha_S^2 \alpha}{12 D(h)} \sum_{\lambda_1,\lambda_2,\lambda_3,\lambda_4}
\Bigl \{
(h - M_H^2) 2 {\rm Re}[C_g C_{\gamma} X_{\lambda_1\lambda_2\lambda_3} Y_{\lambda_4\lambda_4}  
A_{{\lambda_1}{\lambda_2}{\lambda_3}{\lambda_4}{\lambda_4}}^*]
\nonumber \\ && + 
M_H \Gamma_H 2 {\rm Im}[C_g C_{\gamma} X_{\lambda_1\lambda_2\lambda_3} Y_{\lambda_4\lambda_4}  
A_{{\lambda_1}{\lambda_2}{\lambda_3}{\lambda_4}{\lambda_4}}^* ]
\Bigr \}.
\label{eq:Mint2}
\eeq
In the following, we will neglect the small 
effects from $\Gamma_H$, so that, in eq.~(\ref{eq:genform}),
only eq.~(\ref{eq:MH2}) contributes to $P(h)$ and only eq.~(\ref{eq:Mint2}) contributes to 
$I(h)$. The pure continuum cross-section has additional larger contributions from 
$Q\overline Q \rightarrow \gamma\gamma$ and
$Qg \rightarrow Q\gamma\gamma$, as well as from fake photons. 
Significant progress has been made on computing the diphoton backgrounds
\cite{Binoth:1999qq,Bern:2002jx,Campbell:2011bn,Balazs:2006cc,Catani:2011qz}, but
in experimental practice these are determined by fitting to
sidebands, so the pure continuum is not considered here.

\subsection{$Q \overline Q \rightarrow g\gamma\gamma$\label{sec:qqgAA}}
\setcounter{equation}{0}
\setcounter{figure}{0}
\setcounter{table}{0}
\setcounter{footnote}{1}

Next consider the process
\beq
Q(p_1, \lambda_1, j_1) \>+\> \overline Q(p_2, \lambda_2, j_2)
&\longrightarrow&
g(p_3, \lambda_3, a)
\>+\> 
\gamma(p_4, \lambda_4) 
\>+\> 
\gamma(p_5, \lambda_5) 
,
\label{eq:qqgAA}
\eeq
where $j_1$, $j_2$, and $a$ are $SU(3)$ color indices in the anti-fundamental, fundamental, and 
adjoint representations. (The notation means that there is a quark in the initial state, with 
physical momentum and polarization $-p_1$ and $-\lambda_1$, opposite to the outgoing momentum and polarization, and corresponding to an outgoing anti-quark.)
The Higgs-mediated contribution to this process has matrix element:
\beq
{\cal M}^H &=& \frac{g_3}{\sqrt{2}} [{\bf t}^a]_{j_2}{}^{j_1}
C_g C_\gamma Z_{\lambda_1 \lambda_2 \lambda_3} Y_{\lambda_4 \lambda_5}/\left ( h - M_H^2 + i M_H \Gamma_H \right ),
\eeq
where $C_g$, $C_\gamma$, and $Y_{\lambda_4 \lambda_5}$ are as given in 
eqs.~(\ref{eq:defCg}), (\ref{eq:defCgamma}), (\ref{eq:Ypp}), and (\ref{eq:Ypm}) above,
and
\beq
&&
Z_{-+-} \,=\, -\langle 13 \rangle^2/\langle 12 \rangle,
\qquad\qquad
Z_{+-+} \,=\, [ 13 ]^2/[ 12 ],
\label{eq:defZ1}
\\
&&
Z_{+--} \,=\, \langle 23 \rangle^2/\langle 12 \rangle,
\qquad\qquad
Z_{-++} \,=\, -[ 23 ]^2/[ 12 ],
\\
&& 
Z_{+++} = Z_{++-} = Z_{--+} = Z_{---} = 0,
\label{eq:defZ3}
\eeq
and the generator matrices are normalized according to ${\rm Tr}[{\bf t}^a {\bf t}^b] = \delta^{ab}/2$.
For the continuum processes, the matrix elements can be written as:
\beq
{\cal M}^{\rm cont} &=& 
8 \sqrt{2} \pi e_Q^2 \alpha g_3 [{\bf t}^a]_{j_2}{}^{j_1} 
B_{{\lambda_1}{\lambda_2}{\lambda_3}{\lambda_4}{\lambda_5}}
,
\label{eq:MqbqgAA}
\eeq
where $e_Q = +2/3$ or $-1/3$ is the charge of the quark $Q=u,d,s,c,b$.
Because we are specifically interested in interference with diphotons from the Higgs, 
only the matrix elements with $\lambda_4 = \lambda_5$ need to be considered. 
The continuum amplitude equation (\ref{eq:MqbqgAA}) 
vanishes if
$\lambda_3 = \lambda_4 = \lambda_5$. So, there are only four 
helicity configurations that contribute to the interference. They are:
\beq
&& B_{-+-++} \,=\,
\frac{\langle 12 \rangle \langle 13 \rangle^2}{
       \langle 14 \rangle \langle 15 \rangle \langle 24 \rangle \langle 25 \rangle}
,
\qquad\qquad
B_{+-+--} \,=\, -\frac{[12][13]^2}{[14][15][24][25]}
,
\label{eq:defB1}
\\
&& B_{+--++} \,=\,
-\frac{\langle 12 \rangle \langle 23 \rangle^2}{
       \langle 14 \rangle \langle 15 \rangle \langle 24 \rangle \langle 25 \rangle}
,
\qquad\qquad
B_{-++--} \,=\, \frac{[12][23]^2}{[14][15][24][25]}.
\label{eq:defB2}
\eeq
The spin and color sum/average for the reaction eq.(\ref{eq:qqgAA}) is
\beq
{\textstyle \overline{\sum}} &\equiv& \frac{1}{4} \sum_{\lambda_1,\lambda_2,\lambda_3,\lambda_4,\lambda_5} \frac{1}{N^2}
\sum_{j_i,j_2,a}.
\eeq
It follows that:
\beq
{\textstyle \overline{\sum}}\, |{\cal M}^H|^2 &=&
\frac{8\pi\alpha_S}{9D(h)} |C_g C_\gamma|^2 
(s_{13}^2 + s_{23}^2)/{s_{12}} 
,
\label{eq:Hqqbar}
\\
{\textstyle \overline{\sum}}\, 2 {\rm Re}[{\cal M}^H {\cal M}^{{\rm cont}*} ] &=&
\frac{32\pi^2 e_Q^2 \alpha \alpha_S}{9 D(h)} 
\sum_{\lambda_1 \not= \lambda_2, \lambda_3 \not= \lambda_4} \Bigl \{
(h - M_H^2) \,2 {\rm Re} [C_g C_\gamma Z_{\lambda_1 \lambda_2 \lambda_3} Y_{\lambda_4 \lambda_4} B_{\lambda_1 \lambda_2 \lambda_3\lambda_4 \lambda_4}^* ]
\nonumber \\ &&
+ M_H \Gamma_H \,2 {\rm Im} [C_g C_\gamma Z_{\lambda_1 \lambda_2 \lambda_3} Y_{\lambda_4 \lambda_4} B_{\lambda_1 \lambda_2 \lambda_3\lambda_4 \lambda_4}^* ]
\Bigr \} .
\label{eq:intqqbar}
\eeq
The contribution to $pp \rightarrow j\gamma\gamma$ from $Q\overline Q \rightarrow g\gamma\gamma$ 
involving the Higgs is numerically quite small, but is nevertheless included below. More importantly, 
it is useful because it is related by crossing to the processes of the next subsection.

\subsection{$Q g \rightarrow Q\gamma\gamma$ and 
$g\overline Q\rightarrow \overline Q\gamma\gamma$\label{sec:qgqAA}}
\setcounter{equation}{0}
\setcounter{figure}{0}
\setcounter{table}{0}
\setcounter{footnote}{1}

Next consider the process
\beq
Q(p_1, \lambda_1, j_1) \>+\> g(p_2, \lambda_2, a)
&\longrightarrow&
Q(p_3, \lambda_3, j_3)
\>+\> 
\gamma(p_4, \lambda_4) 
\>+\> 
\gamma(p_5, \lambda_5) 
,
\label{eq:qgqAA}
\eeq
The cross-section can be obtained by crossing from the results of the previous section, by 
making the exchange $2 \leftrightarrow 3$ in the spinor helicities 
$\langle ij \rangle$ and $[ij]$ and in $s_{ij}$ in eqs.~(\ref{eq:defZ1})-(\ref{eq:defZ3}) and 
(\ref{eq:defB1})-(\ref{eq:defB2}) and (\ref{eq:Hqqbar}), and multiplying the right sides of
eqs.~(\ref{eq:Hqqbar}) and (\ref{eq:intqqbar}) by $3/8$ to take into account
\beq
{\textstyle \overline{\sum}} &\equiv& \frac{1}{4} \sum_{\lambda_1,\lambda_2,\lambda_3,\lambda_4,\lambda_5} \frac{1}{N(N^2-1)}
\sum_{j_i,j_3,a}.
\eeq
The cross-section for $g\overline Q \rightarrow \overline Q\gamma\gamma$ is obtained in the same way,
except making instead the exchange $1 \leftrightarrow 3$; it gives the same result after integrating over the final state angular variables.

\section{Numerical results\label{sec:num}}
\setcounter{equation}{0}
\setcounter{figure}{0}
\setcounter{table}{0}
\setcounter{footnote}{1}

This section contains numerical results for the shift in the diphoton mass distribution,
as a function of the transverse momentum requirement on the final-state jet:
\beq
p^j_T > p^j_{T,\rm cut} .
\label{eq:pTjetcut}
\eeq
In the numerical integration, this and other cuts are imposed at parton level. 
Equation (\ref{eq:pTjetcut}) is implemented simply by restricting the integrations over
$\hat s$ (or $\tau$) and $\chi$, for fixed $h$,
to the regions
\beq
\hat s &>& h + 2 (p^j_{T,\rm cut})^2 \left [1 + \sqrt{1 + h/(p^j_{T,\rm cut})^2} \right ],
\\
|2 \chi -1 | &<& \sqrt{1 - 4 (p^j_{T,\rm cut})^2/[\hat s (1 - h/\hat s)^2]}.
\eeq
The other cuts are fixed, and implemented within a Monte Carlo integration. 
The jet is required to be central:
\beq
|\eta_j| < 3.0.
\eeq
For the photons, the cuts are
\beq
&&
p^\gamma_{T} \mbox{{(leading, sub-leading)}} > (40, 30)\>{\rm GeV},
\\
&& 
|\eta_{\gamma}| < 2.5,
\\
&&
\Delta R_{\gamma\gamma}, \Delta R_{j\gamma} > 0.4,
\eeq
using the standard definition $\Delta R = \sqrt{(\Delta \eta)^2 + (\Delta \phi)^2}$. 
As in section \ref{sec:H},
I use $M_H = 125$ GeV and $\Gamma_H = 4.2$ MeV, and 
MSTW 2008 NLO \cite{Martin:2009iq} parton distribution functions evaluated at $\mu_F = M_H$, with 
$\alpha_S(\mu_R = 125\>{\rm GeV}) = 0.114629$, and other parameters listed there.

\begin{figure}[!t]
\begin{minipage}[]{0.49\linewidth}
\begin{flushleft}
\includegraphics[width=0.95\linewidth,angle=0]{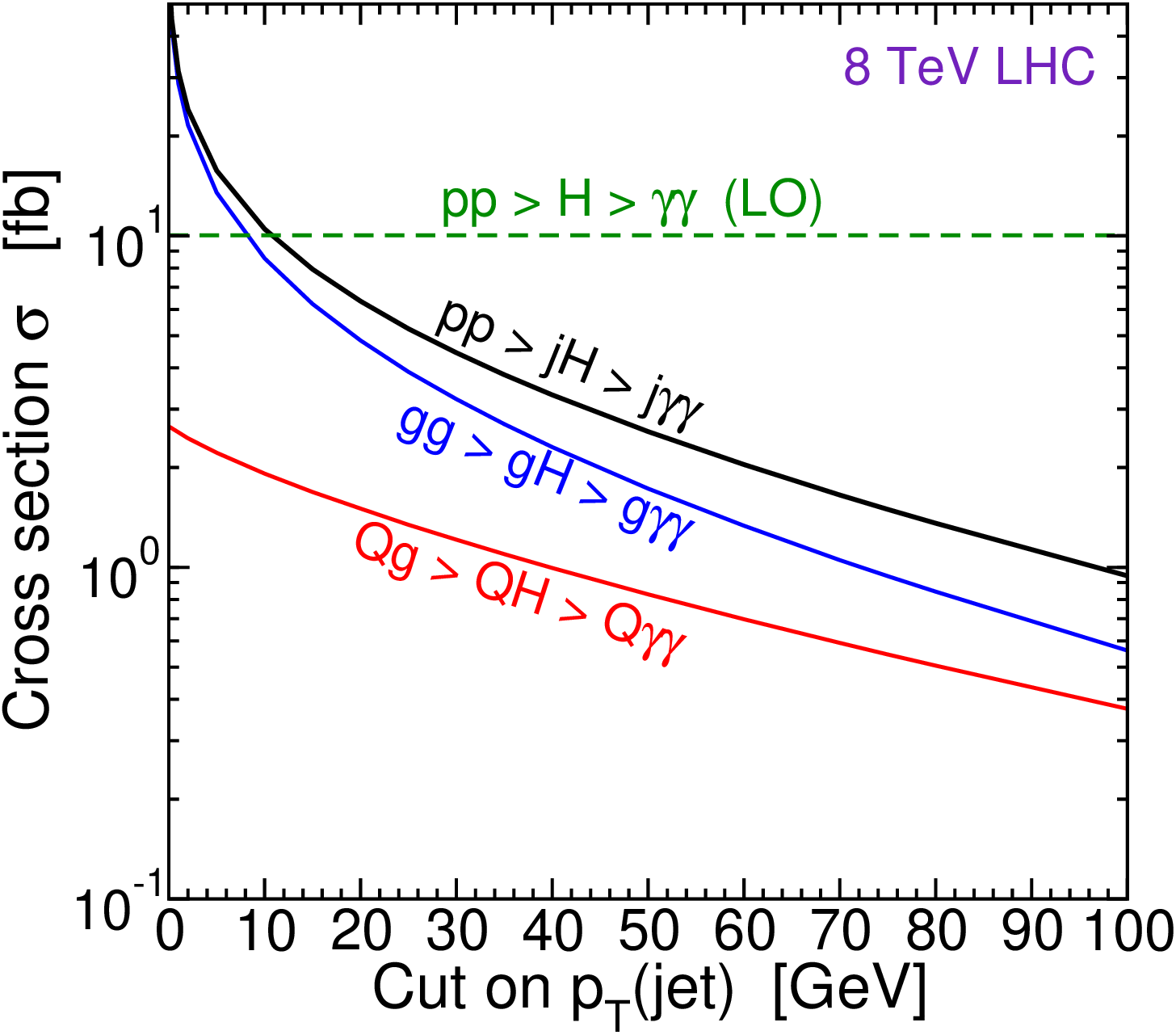}
\end{flushleft}
\end{minipage}
\begin{minipage}[]{0.49\linewidth}
\begin{flushright}
\includegraphics[width=0.95\linewidth,angle=0]{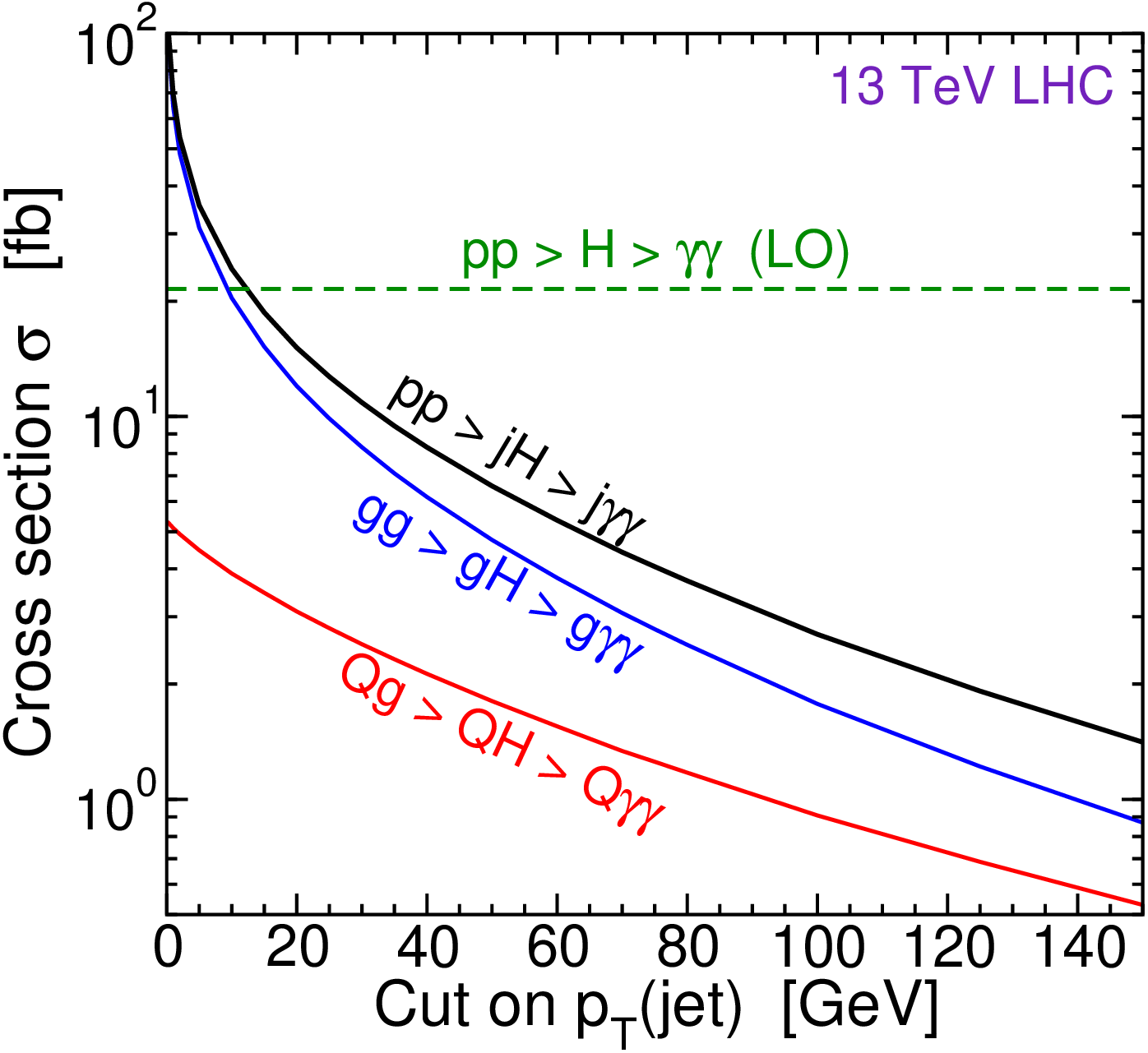}
\end{flushright}
\end{minipage}
\caption{\label{fig:cross} Cross sections for $pp \rightarrow jH \rightarrow j\gamma\gamma$,
as a function of the cut on the transverse momentum of the jet, $p^j_{T,{\rm cut}}$, with other 
cuts and input parameters 
as described in the text. The portions coming from the $gg$ and the $Qg$ (plus $\overline Qg$) 
parton level processes are shown separately; the $Q\overline Q$-initiated process is too small 
to show up on this scale. The cross section level for the leading order $pp \rightarrow H \rightarrow 
\gamma\gamma$ with no jet requirement is also shown as the dashed line. The computation uses MSTW 2008 
NLO parton distribution functions and $\alpha_S$, with $\mu_R = \mu_F = M_H$. 
The left panel is for $pp$ collisions with $\sqrt{s} = 8$ TeV, and the right panel for $\sqrt{s} = 13 $ TeV.} 
\end{figure}

The cross sections for $pp \rightarrow jH \rightarrow j\gamma\gamma$ and its parton-level 
constituents, as a function of the cut
$p^j_{T,{\rm cut}}$, are shown in Figure \ref{fig:cross} for $pp$ collisions with $\sqrt{s} = 8$ 
and 13 TeV. Also shown for comparison is the
cross section for the leading order $pp \rightarrow H \rightarrow 
\gamma\gamma$ with no jet requirement.
The largest contribution is from $gg \rightarrow g H \rightarrow g\gamma\gamma$, especially for small
$p^j_{T,{\rm cut}}$. As expected, that parton-level 
cross section diverges as $p^j_{T,{\rm cut}}$ is taken to 0. The calculated
$pp \rightarrow j H \rightarrow j\gamma\gamma$ cross section exceeds that of the 
leading order tree-level cross section for $pp \rightarrow H \rightarrow \gamma\gamma$, 
with the same cuts on photons,
when $p^j_{T,{\rm cut}} < 10$ GeV (at the 8 TeV LHC) or when $p^j_{T,{\rm cut}} < 12$ GeV 
(at the 13 TeV LHC). The calculation is only physically realistic for larger 
$p^j_{T,{\rm cut}}$, e.g.~30 GeV as in 
\cite{Abdullin:1998er,deFlorian:1999zd,Zmushko:2002fva,Demidov:2004qt,Brein:2007da}.
The partonic processes 
$Qg \rightarrow QH \rightarrow Q\gamma\gamma$ and
$\overline Qg \rightarrow \overline QH \rightarrow \overline Q\gamma\gamma$
(combined in the figure) are subdominant, but certainly not negligible,
while the process $Q\overline Q \rightarrow  gH \rightarrow g\gamma\gamma$
is an order of magnitude below the lower scale of the figure in each case. 

\begin{figure}[!t]
\begin{minipage}[]{0.49\linewidth}
\includegraphics[width=8.2cm,angle=0]{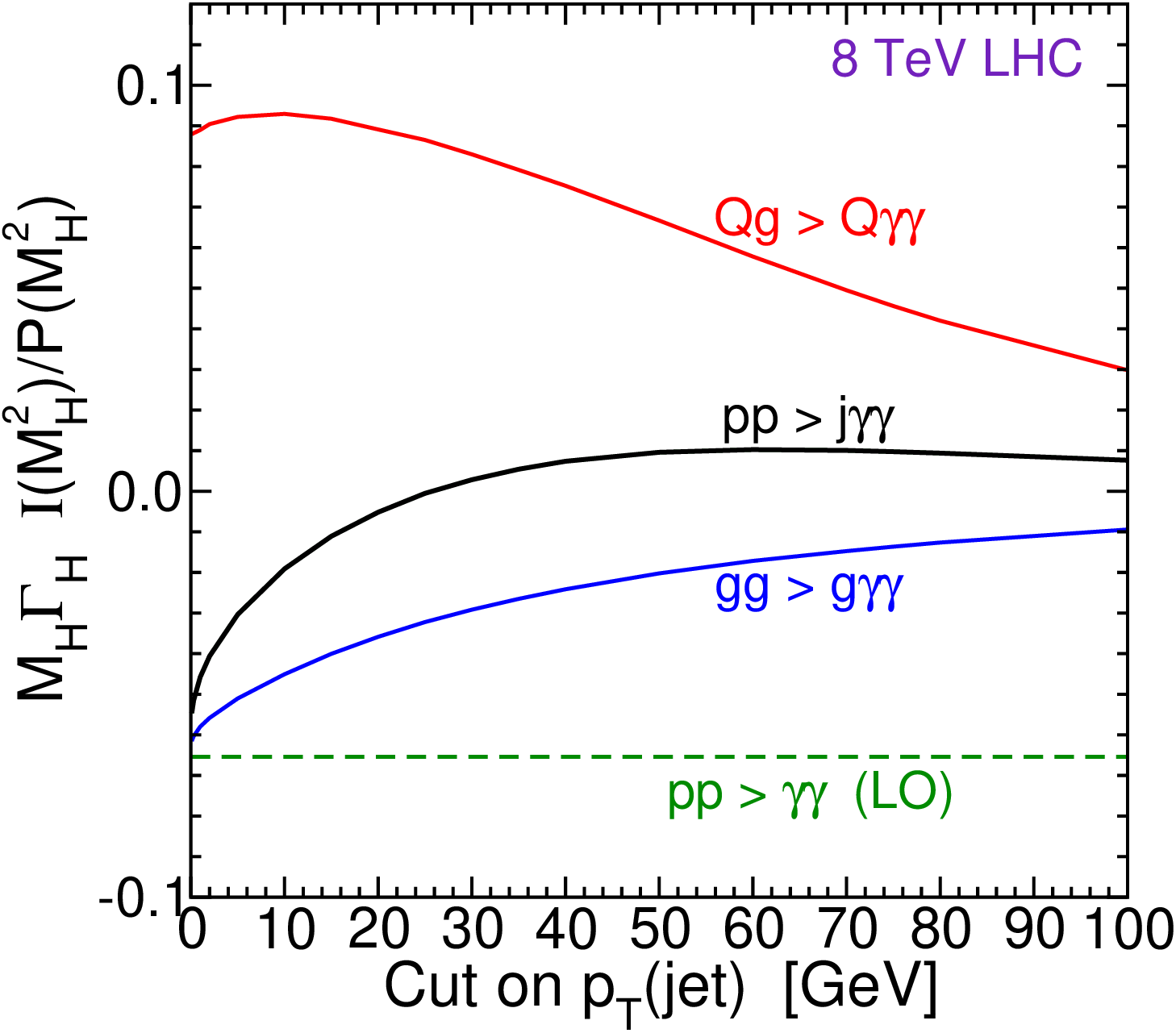}
\end{minipage}
\begin{minipage}[]{0.4\linewidth}
\phantom{x}
\end{minipage}
\begin{minipage}[]{0.45\linewidth}
\caption{\label{fig:IPrat} The quantity $M_H \Gamma_H I(M_H^2)/P(M_H^2)$, where $I(h)$ and $P(h)$
are the functions defined in eq.~(\ref{eq:genform}),
as a function of the cut on the transverse momentum of the jet, $p^j_{T,{\rm cut}}$, with other 
cuts as described in the text, for $gg \rightarrow g\gamma\gamma$, for 
$Qg \rightarrow Q\gamma\gamma$ plus $\overline Qg \rightarrow \overline Q\gamma\gamma$,
and for the combined $pp \rightarrow j \gamma\gamma$, for $\sqrt{s} = 8$ TeV. 
The result for the leading order $pp \rightarrow H \rightarrow \gamma\gamma$ with no jet 
requirement is also shown as the dashed line. The results for $\sqrt{s} = 13$ 
TeV are similar.} 
\end{minipage}
\end{figure}

A simple theoretical measure 
of the relative importance for $\Delta M_{\gamma\gamma}$ 
of the interference compared to the pure resonance
contribution, independent of the details of experimental mass resolution, 
is given by the dimensionless quantity
\beq
M_H \Gamma_H I(M_H^2)/P(M_H^2),
\label{eq:defIPrat}
\eeq
with $I(h)$ and $P(h)$ as defined in eq.~(\ref{eq:genform}).
Equation (\ref{eq:defIPrat}) is
half of the ratio of the maximum deviation from 0 of the unsmeared 
interference lineshape (which occurs at $\sqrt{h}  \approx M_H \pm\Gamma_H/2$) 
compared to the maximum of the pure resonant lineshape (which occurs at $\sqrt{h} = M_H$).
The mass shift $\Delta M_{\gamma\gamma}$ will be approximately proportional to 
eq.~(\ref{eq:defIPrat}), 
with a constant of proportionality that depends 
on mass resolution and other experimental realities including the method used to fit to the data.
The value of this ratio is shown in Figure \ref{fig:IPrat} for 
$gg \rightarrow g\gamma\gamma$, and for 
$Qg \rightarrow Q\gamma\gamma$ combined with $\overline Qg \rightarrow \overline Q\gamma\gamma$,
and for the combined $pp \rightarrow j \gamma\gamma$, for $\sqrt{s} = 8$ TeV.
Note that as $p^j_{T,{\rm cut}}$ approaches 0 (the figure shows the computed values down to 
$p^j_{T,{\rm cut}} = 0.1$ GeV), the result for $gg \rightarrow g\gamma\gamma$ is 
dominated by the log-enhanced contribution from diagrams with an extra gluon attached 
to the $gg \rightarrow \gamma\gamma$ diagrams, and so the
ratio approaches that for the leading order $pp \rightarrow \gamma\gamma$, which is also shown in the 
figure for comparison. For larger $p^j_{T,{\rm cut}}$, the interference contribution 
for the $gg\rightarrow g\gamma\gamma$ process maintains the same (negative) sign but
becomes relatively smaller.
Furthermore, as was already recently found in ref.~\cite{deFlorianHj}, the sign of $I(h)$ is positive
for the $Qg$-initiated process. (The $pp \rightarrow j \gamma\gamma$ curve is 
not the arithmetic sum of the $gg \rightarrow g\gamma\gamma$ and $Qg \rightarrow Q\gamma\gamma$ curves,
because they have different weights in the combination.)
Both of these effects contribute to the fact that 
the interference effect becomes much less important for finite $p^j_{T,{\rm cut}}$, 
 as compared to the 
leading order $pp \rightarrow \gamma\gamma$ process with no jet,
and it has 
the opposite sign for $ p^j_{T,{\rm cut}} > 25$ GeV. The results shown are for
$\sqrt{s} = 8$ TeV; those for $\sqrt{s} = 13$ TeV are quite similar.

\begin{figure}[!t]
\begin{minipage}[]{0.49\linewidth}
\includegraphics[width=8.2cm,angle=0]{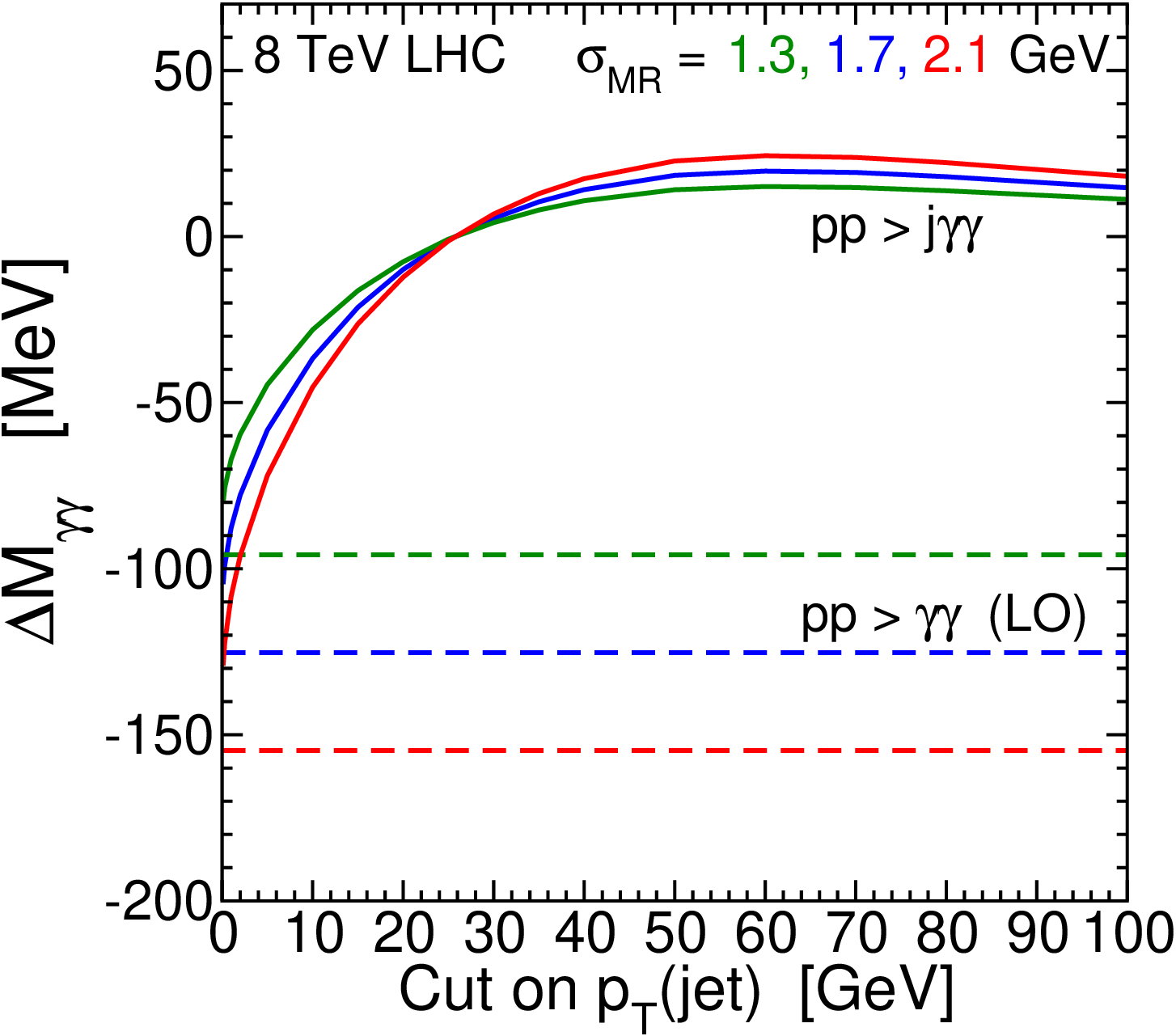}
\end{minipage}
\begin{minipage}[]{0.49\linewidth}
\begin{flushright}
\includegraphics[width=8.2cm,angle=0]{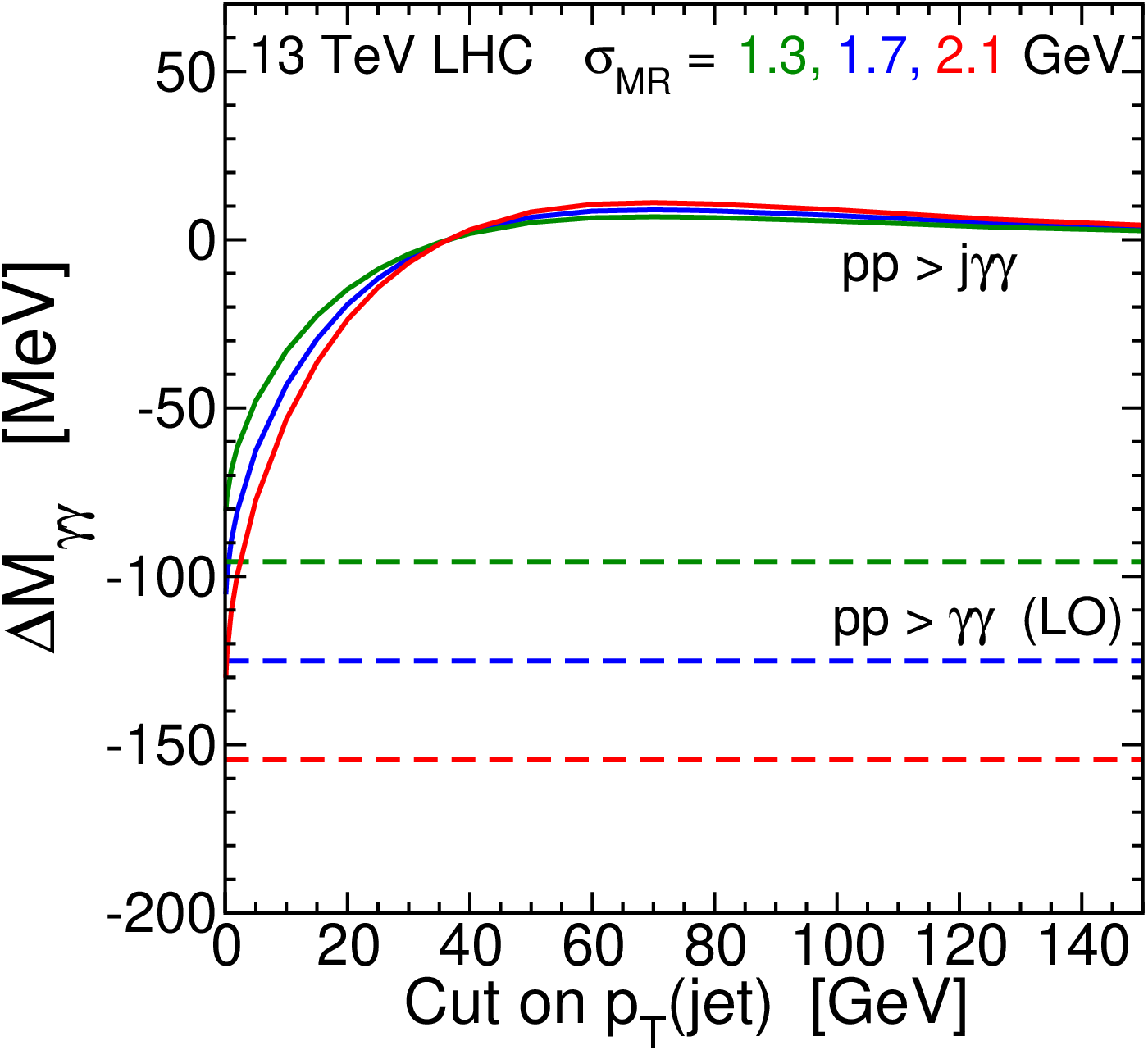}
\end{flushright}
\end{minipage}
\caption{\label{fig:shift} The solid lines show the shifts in the diphoton mass peak, 
$\Delta M_{\gamma\gamma} \equiv M_{\gamma\gamma}^{\rm peak} - M_H$,
for $pp \rightarrow j\gamma\gamma$, as a function of the cut on the transverse 
momentum of the jet, $p^j_{T,{\rm cut}}$, with other 
cuts as described in the text, 
for $\sigma_{\rm MR} = 1.3$, 1.7, and 2.1 GeV (from top to bottom on the left). The dashed lines shows the results for $pp \rightarrow \gamma\gamma$ at leading order without a jet requirement,
again for $\sigma_{\rm MR} = 1.3$, 1.7, and 2.1 GeV (from top to bottom).
The left panel is for $pp$ collisions with
$\sqrt{s} = 8$ TeV, and the right panel for $\sqrt{s} = 13 $ TeV.
} 
\end{figure}

The resulting shifts in the diphoton invariant mass peak,
$\Delta M_{\gamma\gamma} \equiv M_{\gamma\gamma}^{\rm peak} - M_H$, 
for both $\sqrt{s} = 8$ and 13 TeV
are shown in Figure \ref{fig:shift}. These are computed as described in section \ref{sec:H},
using representative Gaussian mass 
resolutions $\sigma_{\rm MR} = 1.3$, 1.7, and 2.1 GeV. (Somewhat larger or smaller shift 
magnitudes could occur for different methods of fitting the lineshape.)
For any reasonable cut $p^j_{T,{\rm cut}}$, the magnitude of the mass shift is much less than 
100 MeV, and it is slightly positive for $p^j_{T,{\rm cut}} > 25$ GeV at the 8 TeV LHC and for
$p^j_{T,{\rm cut}} > 36$ GeV at the 13 TeV LHC. This is in contrast to the negative shift
of about $(-95,-125,-155)$ MeV for $\sigma_{\rm MR} = (1.3, 1.7, 2.4)$ GeV 
from the leading order $pp \rightarrow \gamma\gamma$ case with no jet as found in
section \ref{sec:H} and
\cite{hint}. 

As in the leading order $pp \rightarrow \gamma\gamma$ calculation of section \ref{sec:H}, the 
choices of how to deal with parton distribution functions and $\alpha_S$ and
scale dependences on $\mu_R$ and $\mu_F$
have a big effect on the individual differential cross sections, but these 
tend to cancel out of the mass shift $\Delta M_{\gamma\gamma}$. 
In the case of $gg \rightarrow g\gamma\gamma$, 
both $I(h)$ and $P(h)$ are proportional to $\alpha_S^3$, so this common dependence leads to only a 
small effect on $\Delta M_{\gamma\gamma}$ from choosing between 
NLO or LO $\alpha_S$ or varying $\mu_R$. 
However, for the parton-level processes involving quarks, the function $I(h)$ is proportional to 
$\alpha_S^2$, while $P(h)$ is proportional to $\alpha_S^3$. 
This means that a choice of using the larger LO MSTW 2008 $\alpha_S(M_H)$
would yield a 15\% smaller contribution to the part of the
shift that comes from $Qg \rightarrow Q\gamma\gamma$. A similar effect follows from any 
other renormalization scale choice that uses larger $\alpha_S$ values.
This will tend to shift the predicted value for the total $\Delta M_{\gamma\gamma}$ down 
slightly from the curves shown in Figure \ref{fig:shift}, without changing the conclusion that for 
reasonable values of $p^j_{T,{\rm cut}} $ the magnitude of the shift will be quite small.

\section{Conclusions\label{sec:outlook}}
\setcounter{equation}{0}
\setcounter{figure}{0}
\setcounter{table}{0}
\setcounter{footnote}{1}

In this paper, I have evaluated the shift in the Higgs diphoton mass distribution for 
$pp\rightarrow j\gamma\gamma$ due to interference between the resonant signal and continuum background. 
Unlike the result found in ref.~\cite{hint} at leading order for $pp \rightarrow \gamma\gamma$ 
with no jet, the shift in the mass distribution is probably negligible, less than 
20 MeV in magnitude for $\sigma_{\rm MR} = 1.7$ MeV, when the cut
on the jet transverse momentum is large enough to be realistic. This is due in part to a 
reduction in the relative importance 
of the interference for $gg \rightarrow g\gamma\gamma$ 
as compared to $gg \rightarrow \gamma\gamma$, and in 
part due to the opposite sign of the interference shift
from the $Qg \rightarrow Q\gamma\gamma$ process. The results for
vector boson fusion $pp \rightarrow jj\gamma\gamma$ and the 4-lepton $pp \rightarrow ZZ^*$ final state
should both have very small interference effects. It is therefore tempting to 
speculate that if and
when the Higgs diphoton mass measurement reaches the 100 MeV level of accuracy or better, 
the diphoton mass shift will be appreciable only for the exclusive $pp \rightarrow \gamma\gamma$ 
channel with no additional jets passing cuts like the ones above, 
compared to the other classes of events 
contributing to the mass determination. However, from the results of section \ref{sec:num}, it 
appears that the difference between the diphoton mass peaks for events with no additional 
jets (corresponding to the leading order calculation) 
and those with a central jet with transverse momentum greater than 30 GeV might be as large 
as 150 MeV, for $\sigma_{\rm MR} = 1.7$ MeV. A full calculation including interference 
at NLO, at least, 
for the diphoton mass lineshape would appear to be necessary to make a more
definitive evaluation of this.

{\it Acknowledgments:} 
I am indebted to Daniel de Florian, Lance Dixon, and David Kosower for helpful discussions.
This work was supported in part by the National Science Foundation grant 
number PHY-1068369. 
 


\begin{thebibliography}{90}
\baselineskip=12.6pt

\bibitem{ATLASHiggs}
G.~Aad {\it et al.}  [ATLAS Collaboration],
  ``Observation of a new particle in the search for the Standard Model Higgs boson with the ATLAS detector at the LHC,''
  arXiv:1207.7214 [hep-ex],

\bibitem{CMSHiggs}
  S.~Chatrchyan {\it et al.}  [CMS Collaboration],
  ``Observation of a new boson at a mass of 125 GeV with the CMS experiment at the LHC,''
  Phys.\ Lett.\ B {\bf 716}, 30 (2012)
  [arXiv:1207.7235 [hep-ex]].

\bibitem{ATLAScombination}
  [ATLAS Collaboration],
  ``Combined measurements of the mass and signal strength of the Higgs-like boson with the ATLAS detector using up to 25 fb$^{-1}$ of 
  proton-proton collision data,''
  ATLAS-CONF-2013-014, March 6, 2013.
 
\bibitem{CMScombination} 
  [CMS Collaboration],
  ``Combination of standard model Higgs boson searches and measurements of the properties of the new boson with a mass near 125 GeV''
  CMS-PAS-HIG-12-045, November 16, 2012.
  
\bibitem{Georgi:1977gs}
  H.~M.~Georgi, S.~L.~Glashow, M.~E.~Machacek and D.~V.~Nanopoulos,
  ``Higgs Bosons from Two Gluon Annihilation in Proton Proton Collisions,''
  Phys.\ Rev.\ Lett.\  {\bf 40}, 692 (1978).

\bibitem{Dawson:1990zj}
  S.~Dawson,
  ``Radiative corrections to Higgs boson production,''
  Nucl.\ Phys.\ B {\bf 359}, 283 (1991).

\bibitem{Djouadi:1991tka}
  A.~Djouadi, M.~Spira and P.~M.~Zerwas,
  ``Production of Higgs bosons in proton colliders: QCD corrections,''
  Phys.\ Lett.\ B {\bf 264}, 440 (1991).

\bibitem{Spira:1995rr}
  M.~Spira, A.~Djouadi, D.~Graudenz and P.~M.~Zerwas,
  ``Higgs boson production at the LHC,''
  Nucl.\ Phys.\ B {\bf 453}, 17 (1995)
  [hep-ph/9504378].

\bibitem{deFlorian:1999zd} 
  D.~de Florian, M.~Grazzini and Z.~Kunszt,
  ``Higgs production with large transverse momentum in hadronic collisions at next-to-leading order,''
  Phys.\ Rev.\ Lett.\  {\bf 82}, 5209 (1999)
  [hep-ph/9902483].

\bibitem{Ravindran:2002dc} 
  V.~Ravindran, J.~Smith and W.~L.~Van Neerven,
  ``Next-to-leading order QCD corrections to differential distributions of Higgs boson production in hadron hadron collisions,''
  Nucl.\ Phys.\ B {\bf 634}, 247 (2002)
  [hep-ph/0201114].
  
\bibitem{Glosser:2002gm} 
  C.~J.~Glosser and C.~R.~Schmidt,
  ``Next-to-leading corrections to the Higgs boson transverse momentum spectrum in gluon fusion,''
  JHEP {\bf 0212}, 016 (2002)
  [hep-ph/0209248].

\bibitem{Campbell:2006xx} 
  J.~M.~Campbell, R.~K.~Ellis and G.~Zanderighi,
  ``Next-to-Leading order Higgs + 2 jet production via gluon fusion,''
  JHEP {\bf 0610}, 028 (2006)
  [hep-ph/0608194].
  
\bibitem{Campbell:2010cz} 
  J.~M.~Campbell, R.~K.~Ellis and C.~Williams,
  ``Hadronic production of a Higgs boson and two jets at next-to-leading order,''
  Phys.\ Rev.\ D {\bf 81}, 074023 (2010)
  [arXiv:1001.4495 [hep-ph]].

\bibitem{vanDeurzen:2013rv} 
  H.~van Deurzen, 
  {\it et al.},
  ``NLO QCD corrections to the production of Higgs plus two jets at the LHC,''
  arXiv:1301.0493 [hep-ph].

\bibitem{Harlander:2002wh}
  R.~V.~Harlander and W.~B.~Kilgore,
  ``Next-to-next-to-leading order Higgs production at hadron colliders,''
  Phys.\ Rev.\ Lett.\  {\bf 88}, 201801 (2002)
  [hep-ph/0201206].

\bibitem{Anastasiou:2002yz}
  C.~Anastasiou and K.~Melnikov,
  ``Higgs boson production at hadron colliders in NNLO QCD,''
  Nucl.\ Phys.\ B {\bf 646}, 220 (2002)
  [hep-ph/0207004].

\bibitem{Ravindran:2003um}
  V.~Ravindran, J.~Smith and W.~L.~van Neerven,
  ``NNLO corrections to the total cross-section for Higgs boson production in hadron hadron collisions,''
  Nucl.\ Phys.\ B {\bf 665}, 325 (2003)
  [hep-ph/0302135].

\bibitem{Anastasiou:2005qj}
  C.~Anastasiou, K.~Melnikov and F.~Petriello,
  ``Fully differential Higgs boson production and the di-photon signal through next-to-next-to-leading order,''
  Nucl.\ Phys.\ B {\bf 724}, 197 (2005)
  [hep-ph/0501130].

\bibitem{Boughezal:2013uia} 
  R.~Boughezal, F.~Caola, K.~Melnikov, F.~Petriello and M.~Schulze,
  ``Higgs boson production in association with a jet at next-to-next-to-leading order 
  in perturbative QCD,''
  arXiv:1302.6216 [hep-ph].

\bibitem{Aglietti:2004nj}
  U.~Aglietti, R.~Bonciani, G.~Degrassi and A.~Vicini,
  ``Two loop light fermion contribution to Higgs production and decays,''
  Phys.\ Lett.\ B {\bf 595}, 432 (2004)
  [hep-ph/0404071].

\bibitem{Actis:2008ug}
  S.~Actis, G.~Passarino, C.~Sturm and S.~Uccirati,
  ``NLO Electroweak Corrections to Higgs Boson Production at Hadron Colliders,''
  Phys.\ Lett.\ B {\bf 670}, 12 (2008)
  [arXiv:0809.1301 [hep-ph]].

\bibitem{Anastasiou:2008tj}
  C.~Anastasiou, R.~Boughezal and F.~Petriello,
  ``Mixed QCD-electroweak corrections to Higgs boson production in gluon fusion,''
  JHEP {\bf 0904}, 003 (2009)
  [arXiv:0811.3458 [hep-ph]].

\bibitem{Catani:2003zt}
  S.~Catani, D.~de Florian, M.~Grazzini and P.~Nason,
  ``Soft gluon resummation for Higgs boson production at hadron colliders,''
  JHEP {\bf 0307}, 028 (2003)
  [hep-ph/0306211].

\bibitem{Bozzi:2005wk} 
  G.~Bozzi, S.~Catani, D.~de Florian and M.~Grazzini,
  ``Transverse-momentum resummation and the spectrum of the Higgs boson at the LHC,''
  Nucl.\ Phys.\ B {\bf 737}, 73 (2006)
  [hep-ph/0508068].

\bibitem{deFlorian:2011xf}
  D.~de Florian, G.~Ferrera, M.~Grazzini and D.~Tommasini,
  ``Transverse-momentum resummation: Higgs boson production at the Tevatron and the LHC,''
  JHEP {\bf 1111}, 064 (2011)
  [arXiv:1109.2109 [hep-ph]].

\bibitem{deFlorian:2009hc}
  D.~de Florian and M.~Grazzini,
  ``Higgs production through gluon fusion: Updated cross sections at the Tevatron and the LHC,''
  Phys.\ Lett.\ B {\bf 674}, 291 (2009)
  [arXiv:0901.2427 [hep-ph]].

\bibitem{Dittmaier:2011ti}
  S.~Dittmaier {\it et al.}  [LHC Higgs Cross Section Working Group Collaboration],
  ``Handbook of LHC Higgs Cross Sections: 1. Inclusive Observables,''
  arXiv:1101.0593 [hep-ph].

\bibitem{Dittmaier:2012vm}
  S.~Dittmaier, {\it et al.},
  ``Handbook of LHC Higgs Cross Sections: 2. Differential Distributions,''
  arXiv:1201.3084 [hep-ph].

\bibitem{Anastasiou:2012hx}
  C.~Anastasiou, S.~Buehler, F.~Herzog and A.~Lazopoulos,
  ``Inclusive Higgs boson cross-section for the LHC at 8 TeV,''
  JHEP {\bf 1204}, 004 (2012)
  [arXiv:1202.3638 [hep-ph]].

\bibitem{hint}
S.~P.~Martin,
  ``Shift in the LHC Higgs diphoton mass peak from interference with background,''
  Phys.\ Rev.\ D {\bf 86}, 073016 (2012)
  [arXiv:1208.1533 [hep-ph]].

\bibitem{Ellis:1975ap}
  J.~R.~Ellis, M.~K.~Gaillard and D.~V.~Nanopoulos,
  ``A Phenomenological Profile of the Higgs Boson,''
  Nucl.\ Phys.\ B {\bf 106}, 292 (1976).

\bibitem{Shifman:1979eb}
  M.~A.~Shifman, A.~I.~Vainshtein, M.~B.~Voloshin and V.~I.~Zakharov,
  ``Low-Energy Theorems for Higgs Boson Couplings to Photons,''
  Sov.\ J.\ Nucl.\ Phys.\  {\bf 30}, 711 (1979)
  [Yad.\ Fiz.\  {\bf 30}, 1368 (1979)].

\bibitem{Gunion:1985dj}
  J.~F.~Gunion, P.~Kalyniak, M.~Soldate and P.~Galison,
  ``Searching For The Intermediate Mass Higgs Boson,''
  Phys.\ Rev.\ D {\bf 34}, 101 (1986).

\bibitem{Ellis:1987xu}
  R.~K.~Ellis, I.~Hinchliffe, M.~Soldate and J.~J.~van der Bij,
  ``Higgs Decay to tau+ tau-: A Possible Signature of Intermediate Mass Higgs Bosons at the SSC,''
  Nucl.\ Phys.\ B {\bf 297}, 221 (1988).

\bibitem{Gunion:1987ke}
  J.~F.~Gunion, G.~L.~Kane and J.~Wudka,
  ``Search Techniques for Charged and Neutral Intermediate Mass Higgs Bosons,''
  Nucl.\ Phys.\ B {\bf 299}, 231 (1988).

\bibitem{Djouadi:1997yw}
  A.~Djouadi, J.~Kalinowski and M.~Spira,
  ``HDECAY: A Program for Higgs boson decays in the standard model and its supersymmetric extension,''
  Comput.\ Phys.\ Commun.\  {\bf 108}, 56 (1998)
  [hep-ph/9704448].

\bibitem{Dicus:1987fk} 
  D.A.~Dicus and S.S.D.~Willenbrock,
  ``Photon Pair Production And The Intermediate Mass Higgs Boson,''
  Phys.\ Rev.\ D {\bf 37}, 1801 (1988).

\bibitem{Dixon:2003yb} 
  L.J.~Dixon and M.S.~Siu,
  ``Resonance continuum interference in the diphoton Higgs signal at the LHC,''
  Phys.\ Rev.\ Lett.\  {\bf 90}, 252001 (2003)
  [hep-ph/0302233].

\bibitem{Bern:2001df}
  Z.~Bern, A.~De Freitas and L.J.~Dixon,
  ``Two loop amplitudes for gluon fusion into two photons,''
  JHEP {\bf 0109}, 037 (2001)
  [hep-ph/0109078].

\bibitem{Glover:1988fe}
  E.~W.~N.~Glover and J.~J.~van der Bij,
  ``Vector Boson Pair Production Via Gluon Fusion,''
  Phys.\ Lett.\ B {\bf 219}, 488 (1989).
%
  E.~W.~N.~Glover and J.~J.~van der Bij,
  ``Z Boson Pair Production Via Gluon Fusion,''
  Nucl.\ Phys.\ B {\bf 321}, 561 (1989).

\bibitem{Dixon:2008xc}
  L.J.~Dixon and Y.~Sofianatos,
  ``Resonance-Continuum Interference in Light Higgs Boson Production at a Photon Collider,''
  Phys.\ Rev.\ D {\bf 79}, 033002 (2009)
  [arXiv:0812.3712 [hep-ph]].

\bibitem{Campbell:2011cu} 
  J.~M.~Campbell, R.~K.~Ellis and C.~Williams,
  ``Gluon-Gluon Contributions to W+ W- Production and Higgs Interference Effects,''
  JHEP {\bf 1110}, 005 (2011)
  [arXiv:1107.5569 [hep-ph]].

\bibitem{Passarino:2012ri}
  G.~Passarino,
  ``Higgs Interference Effects in $gg \rightarrow ZZ$ and their Uncertainty,''
  arXiv:1206.3824 [hep-ph].

\bibitem{Kauer:2012hd}
  N.~Kauer and G.~Passarino,
  ``Inadequacy of zero-width approximation for a light Higgs boson signal,''
  arXiv:1206.4803 [hep-ph].

\bibitem{deFlorianHj} D.~de Florian, N.~Fidanza, R.~Hern\'andez-Pinto, J.~Mazzitelli, Y.~Rotstein Habarnau, and 
G.~Sborlini, ``A complete ${\cal O}(\alpha_S^2)$ calculation of the signal-background interference for the Higgs 
diphoton decay channel", 
  arXiv:1303.1397 [hep-ph].

\bibitem{Abdullin:1998er} 
  S.~Abdullin, M.~Dubinin, V.~Ilyin, D.~Kovalenko, V.~Savrin and N.~Stepanov,
  ``Higgs boson discovery potential of LHC in the channel $p p \rightarrow \gamma \gamma $+ jet,''
  Phys.\ Lett.\ B {\bf 431}, 410 (1998)
  [hep-ph/9805341].

\bibitem{Zmushko:2002fva} 
  V.~V.~Zmushko,
  ``Search for H $\rightarrow \gamma \gamma$ in association with one jet,''
  ATL-PHYS-2002-020.

\bibitem{Demidov:2004qt} 
  S.~V.~Demidov and D.~S.~Gorbunov,
  ``LHC prospects in searches for neutral scalars in $pp \rightarrow \gamma \gamma$+ jet: SM Higgs boson, radion, sgoldstino,''
  Phys.\ Atom.\ Nucl.\  {\bf 69}, 712 (2006)
  [hep-ph/0405213].

\bibitem{Brein:2007da} 
  O.~Brein and W.~Hollik,
  ``Distributions for MSSM Higgs boson + jet production at hadron colliders,''
  Phys.\ Rev.\ D {\bf 76}, 035002 (2007)
  [arXiv:0705.2744 [hep-ph]].

\bibitem{Kauffman}
R.P.~Kauffman,
  ``Higgs boson p(T) in gluon fusion,''
  Phys.\ Rev.\ D {\bf 44}, 1415 (1991).
S.~Dawson and R.P.~Kauffman,
  ``Higgs boson plus multi - jet rates at the SSC,''
  Phys.\ Rev.\ Lett.\  {\bf 68}, 2273 (1992);
  ``QCD corrections to H $\rightarrow \gamma \gamma$,''
  Phys.\ Rev.\ D {\bf 47}, 1264 (1993);
  ``QCD corrections to Higgs boson production: nonleading terms in the heavy quark limit,''
  Phys.\ Rev.\ D {\bf 49}, 2298 (1994)
  [hep-ph/9310281].
R.P.~Kauffman, S.~V.~Desai and D.~Risal,
  ``Production of a Higgs boson plus two jets in hadronic collisions,''
  Phys.\ Rev.\ D {\bf 55}, 4005 (1997)
  [Erratum-ibid.\ D {\bf 58}, 119901 (1998)]
  [hep-ph/9610541].
  
 \bibitem{Karplus:1950zz}
  R.~Karplus and M.~Neuman,
  ``The scattering of light by light,''
  Phys.\ Rev.\  {\bf 83}, 776 (1951).

\bibitem{Costantini:1971cj}
  V.~Costantini, B.~De Tollis and G.~Pistoni,
  ``Nonlinear effects in quantum electrodynamics,''
  Nuovo Cim.\ A {\bf 2}, 733 (1971).

\bibitem{Combridge:1980sx}
  B.L.~Combridge,
  ``Consequences Of The Photon-Gluon Induced Couplings Of QCD,''
  Nucl.\ Phys.\ B {\bf 174}, 243 (1980).
  
  
\bibitem{Martin:2009iq}
  A.~D.~Martin, W.~J.~Stirling, R.~S.~Thorne and G.~Watt,
  ``Parton distributions for the LHC,''
  Eur.\ Phys.\ J.\ C {\bf 63}, 189 (2009)
  [arXiv:0901.0002 [hep-ph]].
  
\bibitem{ManganoParke}
M.~L.~Mangano and S.~J.~Parke,
  ``Multiparton amplitudes in gauge theories,''
  Phys.\ Rept.\  {\bf 200}, 301 (1991)
  [hep-th/0509223].

\bibitem{DixonTASI}
  L.~J.~Dixon,
  ``Calculating scattering amplitudes efficiently,''
  TASI 1995 lectures,
  [hep-ph/9601359].
  
\bibitem{Bern:1993mq}
Z.~Bern, L.~J.~Dixon and D.~A.~Kosower,
  ``One loop corrections to five gluon amplitudes,''
  Phys.\ Rev.\ Lett.\  {\bf 70}, 2677 (1993)
  [hep-ph/9302280].

\bibitem{deFlorian:1999tp} 
  D.~de Florian and Z.~Kunszt,
  ``Two photons plus jet at LHC: The NNLO contribution from the g g initiated process,''
  Phys.\ Lett.\ B {\bf 460}, 184 (1999)
  [hep-ph/9905283].
  
\bibitem{Balazs:1999yf} 
  C.~Balazs, P.~M.~Nadolsky, C.~Schmidt and C.~P.~Yuan,
  ``Diphoton background to Higgs boson production at the LHC with soft gluon effects,''
  Phys.\ Lett.\ B {\bf 489}, 157 (2000)
  [hep-ph/9905551].

\bibitem{Binoth:1999qq}
  T.~Binoth, J.~P.~Guillet, E.~Pilon and M.~Werlen,
  ``A Full next-to-leading order study of direct photon pair production in hadronic collisions,''
  Eur.\ Phys.\ J.\ C {\bf 16}, 311 (2000)
  [hep-ph/9911340];
``A Next-to-leading order study of photon pion and pion pair hadro production in the light of the Higgs boson search at the LHC,''
  Eur.\ Phys.\ J.\ direct C {\bf 4}, 7 (2002)
  [hep-ph/0203064].

\bibitem{Bern:2002jx}
  Z.~Bern, L.J.~Dixon and C.~Schmidt,
  ``Isolating a light Higgs boson from the diphoton background at the CERN LHC,''
  Phys.\ Rev.\ D {\bf 66}, 074018 (2002)
  [hep-ph/0206194].

\bibitem{Campbell:2011bn}
  J.~M.~Campbell, R.~K.~Ellis and C.~Williams,
  ``Vector boson pair production at the LHC,''
  JHEP {\bf 1107}, 018 (2011)
  [arXiv:1105.0020 [hep-ph]].

\bibitem{Balazs:2006cc}
  C.~Balazs, E.~L.~Berger, P.~M.~Nadolsky and C.~-P.~Yuan,
  ``All-orders resummation for diphoton production at hadron colliders,''
  Phys.\ Lett.\ B {\bf 637}, 235 (2006)
  [hep-ph/0603037].
  P.~M.~Nadolsky, C.~Balazs, E.~L.~Berger and C.~-P.~Yuan,
  ``Gluon-gluon contributions to the production of continuum diphoton pairs at hadron colliders,''
  Phys.\ Rev.\ D {\bf 76}, 013008 (2007)
  [hep-ph/0702003 [HEP-PH]].
  C.~Balazs, E.~L.~Berger, P.~M.~Nadolsky and C.~-P.~Yuan,
  ``Calculation of prompt diphoton production cross-sections at Tevatron and LHC energies,''
  Phys.\ Rev.\ D {\bf 76}, 013009 (2007)
  [arXiv:0704.0001 [hep-ph]].

\bibitem{Catani:2011qz}
  S.~Catani, L.~Cieri, D.~de Florian, G.~Ferrera and M.~Grazzini,
  ``Diphoton production at hadron colliders: a fully-differential QCD
calculation at NNLO,''
  Phys.\ Rev.\ Lett.\  {\bf 108}, 072001 (2012)
  [arXiv:1110.2375 [hep-ph]].


\end{thebibliography}
\end{document}